\documentclass[a4paper,aps,prd, preprint, preprintnumbers, superscriptaddress, nofootinbib]{revtex4-1}
\pdfoutput=1
\usepackage{fullpage}
\usepackage{amsfonts}
\usepackage{amsmath}
\usepackage{slashed}
\usepackage{amssymb}
\usepackage{graphicx}
\usepackage{epic}
\usepackage{mathtools}
\usepackage{eepic}
\usepackage{epsfig}
\usepackage{latexsym}
\usepackage{color}
\usepackage{float}
\usepackage{multirow}
\usepackage{hyperref}
\usepackage[caption=false]{subfig}
\usepackage{natbib}
\usepackage{cancel}
\newcommand{\U}[1]{\mathrm{U}(1)_{\mathrm{#1}}}			
\newcommand{\SU}[2]{\mathrm{SU}(#1)_{\mathrm{#2}}}	

\usepackage{shorthand}
\usepackage{parskip}

\usepackage{mciteplus}

\begin{document}

\begin{flushright}
LU TP 17-36\\
January 2018
\end{flushright}

\title{Heavy charged scalars from $c\bar{s}$ fusion: A generic search strategy applied 
to a 3HDM with $\mathrm{U}(1) \times \mathrm{U}(1)$ family symmetry}

\author{Jos\'e Eliel Camargo-Molina}
\email{eliel@thep.lu.se}
\affiliation{Department of Astronomy and Theoretical Physics, Lund University, SE-223 62 Lund, Sweden}

\author{Tanumoy Mandal}
\email{tanumoy.mandal@physics.uu.se}
\affiliation{Department of Physics and Astronomy, Uppsala University, Box 516, SE-751 20 Uppsala, Sweden}
\affiliation{Department of Physics and Astrophysics, University of Delhi, Delhi 110007, India}

\author{Roman Pasechnik}
\email{roman.pasechnik@thep.lu.se}
\affiliation{Department of Astronomy and Theoretical Physics, Lund University, SE-223 62 Lund, Sweden}

\author{Jonas Wess\'en}
\email{jonas.wessen@thep.lu.se}
\affiliation{Department of Astronomy and Theoretical Physics, Lund University, SE-223 62 Lund, Sweden}


\begin{abstract}
We describe a class of three Higgs doublet models (3HDMs) with a softly broken $\U{} \times \U{}$ family symmetry that enforces a Cabibbo-like quark mixing while 
forbidding tree-level flavour changing neutral currents. The hierarchy in the observed quark masses is partly explained by a softer hierarchy in the vacuum expectation 
values of the three Higgs doublets. As a consequence, the physical scalar spectrum contains a Standard Model (SM) like Higgs boson $h_{125}$ while exotic scalars 
couple the strongest to the second quark family, leading to rather unconventional discovery channels that could be probed at the Large Hadron Collider. In particular, 
we describe a search strategy for the lightest charged Higgs boson $H^\pm$, through the process $c\bar s\to H^{+}\to W^+\,h_{125}$, using a multivariate analysis 
that leads to an excellent discriminatory power against the SM background. Although the analysis is applied to the proposed class of 3HDMs, we employ a 
model-independent formulation such that it can be applied to any other model with the same discovery channel. 
\end{abstract}


\maketitle

\section{Introduction}
\label{sec:intro}

The Standard Model (SM) remarkably stands as one of the most successful theories in physics. However, it can still be considered rather \textit{ad hoc} in its nature, 
with unexplained features that arise from fitting the experimental data. In addition, it fails to offer an explanation to several observed natural phenomena such as 
dark matter, neutrino masses or baryon asymmetry in the universe. It is then natural to study extensions of the SM that, while retaining its predictive power, 
offer explanations or shed light into the origin of e.g.~the hierarchy of fermion masses or rather specific flavour structure of the SM. There is a plethora of such 
beyond the SM (BSM) theories, but not many of those offer unconventional features testable at the current experiments. 

One of the simplest and most studied extensions is the class of the so-called Two-Higgs Doublet Models (2HDMs) that add a second $\SU{2}{L}$ doublet to the SM 
(an extensive review can be found in Ref.~\cite{Branco:2011iw}). The 2HDMs offer interesting phenomenological signatures and can lead to e.g.~extra sources of CP 
violation, dark matter candidates and stable vacua at high energies. However, they typically introduce many new free parameters, fail to address the origin of the mass 
hierarchy in the fermion sector of the SM and require extra discrete symmetries to avoid tree-level Flavour Changing Neutral Currents (FCNCs). 

The Three Higgs Doublet Models (3HDMs) can overcome some of those limitations (see e.g.~Refs.~\cite{PhysRevLett.37.657, BRANCO1985306, BOTELLA2010194}) 
and have sparked interest in recent literature (see e.g.~Refs.~\cite{Ivanov:2014doa, Akeroyd:2016gvk, Camargo-Molina:2017rwj,Merchand:2016ldu,Moretti:2015cwa,
Das:2014fea, Emmanuel-Costa:2016vej}). While retaining most of the features of 2HDMs, 3HDMs can offer explanations to yet unexplained features of the SM with 
predictions testable in the current collider measurements. In particular, the increased field content makes it possible to impose higher symmetries, which in turn can 
lead to interesting flavour structures.

As shown in Refs.~\cite{Keus:2013hya, Ivanov:2011ae}, the most constraining realisable abelian symmetry of the scalar potential in 3HDM is $\U{} \times \U{}$. In this work, we promote 
the $\U{} \times \U{}$ symmetry of the scalar sector to the fermion sector, hereinafter called $\U{X}\times \U{Z}$, in such a way that (1) no tree-level FCNCs are present, 
(2) a Cabibbo-like mixing is enforced, and (3) the fermion mass hierarchies are related to a hierarchy in the three vacuum expectation values (VEVs) of the doublets. 
This leads to a model that, although remarkably simple due to its high symmetry, is still capable of both reproducing the current experimental data and providing 
the exotic collider signatures. The latter is due to the fact that, as a consequence of the model symmetries, the new scalar states (both charged and neutral) couple 
dominantly to the second quark family. 

At the LHC, the searches for charged Higgs bosons are generally categorized into two mass regions depending on whether its mass $m_{H^\pm}$ is smaller 
or bigger than the top quark mass $m_t$. The motivation of this categorization comes from the properties of $H^\pm$ within the various 2HDM types 
or supersymmetric models. Usually, for a heavy charged Higgs state $(m_{H^{\pm}} \gtrsim m_t)$, the dominant production and decay channels in the LHC 
context are $pp\to H^{-}t\bar{b}\,(H^{+}\bar{t}b)$ and $H^+\to t\bar{b}$ ($H^-\to\bar{t}b$), respectively~\cite{ATLAS:2016qiq,Khachatryan:2015qxa}. 
Apart from this channel, production of $H^{\pm}$ in vector boson ($W^\pm Z$) fusion followed by the $H^{\pm}\to W^\pm Z$ decay is prominent 
in Higgs triplet models such as the Georgi-Machacek model~\cite{Georgi:1985nv}. This channel has also been searched for by the ATLAS~\cite{Aad:2015nfa} 
and CMS~\cite{Sirunyan:2017sbn} collaborations recently. Conversely, a light charged Higgs boson $(m_{H^\pm} \lesssim m_t)$ that decays to $\tau\bar{\nu}$~\cite{Aad:2014kga,
Khachatryan:2015qxa}, $c\bar{s}$~\cite{Aad:2013hla,Khachatryan:2015uua} or $c\bar{b}$~\cite{CMS:2016qoa} has also been searched for at the LHC.
Previously, at LEP, pair production of $H^\pm$ was considered where $H^\pm$ subsequently decays to a $W^\pm A$ pair~\cite{Abdallah:2003wd,
Abbiendi:2008aa} (where $A$ is a scalar with mass $m_A>12$ GeV and predominantly decays to $b\bar{b}$ pairs).

Searches for heavy $H^\pm$ become increasingly important with the rise of the LHC center-of-mass energy and luminosity, thus it is important to explore new 
production and decay modes of $H^\pm$ that are predicted by various BSM theories. In this paper, we particularly focus 
on a new search channel where a heavy $H^\pm$ resonantly decays to a $W^\pm h_{125}$ pair after being produced in $c\bar{s}$ ($\bar{c}s$) fusion. 
This rather uncommon search channel leads to testable predictions of our model at current LHC energies. In Refs.~\cite{Moretti:2000yg,Enberg:2014pua,
Moretti:2016jkp}, the $H^\pm\to W^\pm h_{125}$ decay is considered where the $H^-$ ($H^+$) is produced in association with a $t\bar{b}$ ($\bar{t}b$) pair. 
In our case, $H^+$ is produced in the $s$-channel resonance through $c\bar{s}$ fusion. In Ref.~\cite{Dittmaier:2007uw}, the possibility of sizable $c\bar{s}\to H^+$ 
production cross section is discussed in the SUSY context where a squark mixing can circumvent the chiral suppression of the single $H^\pm$ production. 
In Ref.~\cite{Altmannshofer:2016zrn}, $c\bar{s}\to H^+$ production is shown to be dominant for a 2HDM with a Yukawa sector chosen such that one doublet 
couples strongest to the second generation. In our model, we will see that the $c\bar{s}H^+$ and $\bar{c}sH^-$ couplings are sizable due to the hierarchy in 
VEVs combined with the particular structure of the Yukawa sector.

In section~\ref{sec:3HDM}, we introduce the model, its fermion and scalar sectors and their interplay as given by the $\U{X}\times \U{Z}$ symmetry, 
the VEV hierarchy and the spectrum of the theory. In section~\ref{sec:hcprod}, we discuss the charged Higgs boson production and decay channels of the model 
and introduce a model-independent way to study $H^\pm$ production in the same unconventional channels. In section~\ref{sec:ColliderPheno}, we show the results 
of a multivariate analysis for the charged Higgs boson searches and show, by using the results of a genetic algorithm scan, that our proposed theory can produce 
the type of signals visible with such an analysis at the LHC. Finally, we summarize and conclude our results in section~\ref{sec:sumconclu}.

\section{The model} 
\label{sec:3HDM}

In this work, we propose a 3HDM, with features that lead to a simple yet predictive theory. The model has a $\U{X}\times \U{Z}$ global symmetry constraining its scalar potential. This symmetry is the biggest abelian symmetry not leading to additional accidental symmetries in a 3HDM \cite{Ivanov:2014doa, Keus:2013hya, Ivanov:2011ae}. As a consequence, in the limit of one VEV being much larger than the other two, we can derive simple analytical formulas for masses and mixing matrices in the scalar sector and readily understand the features of the model and its physical consequences. 

The $\U{X}\times \U{Z}$ is also present in the fermion sector of the theory. We choose the charge assignments to constrain the Yukawa sector in a manner consistent 
with the experimental hierarchies in the quark mass spectrum while forbidding tree-level FCNCs arising from the scalar sector. The upside of this is that the mass hierarchy 
is directly connected to a VEV hierarchy, which needs not to be as strong as the hierarchy in the SM Yukawa parameters to explain the known quark masses. 

A nice consequence is the opening of new search strategies for testing this model at the LHC. Due to the structure of the Yukawa sector, two physical charged Higgs 
bosons would be produced mainly through $c\bar{s}$ fusion, which can lead to good signal-to-background ratios as will be shown in section~\ref{sec:ColliderPheno}. 

\subsection{VEV hierarchy and the softly broken $\U{X} \times \U{Z}$ symmetry}

Besides the field content of the SM, the model has two additional scalar $\SU{2}{L}$ doublets for a total of three. We will denote them by $H_{1,2,3}$, with charges 
shown in Table~\ref{tab:U1Charges}, and expand around the vacuum as
\begin{equation}
H_i = \begin{pmatrix}
H^+_i \\
\frac{1}{\sqrt{2}}\left(v_i + h_i + \mathrm{i} A_i \right) 
\end{pmatrix} \, , \quad i=1,2,3 \, .
\end{equation}
We will often focus on the case where $v_3 \gg v_{1,2}$. This particular limit calls for the definition of a small parameter $\xi$, 

\begin{equation}
\xi \equiv \frac{\sqrt{v_1^2+v_2^2}}{v_3} \,.
\end{equation}
After spontaneous symmetry breaking, in the limit $\xi \rightarrow 0$, there remains a $\U{X} \times \U{YZ}$ symmetry where $\U{YZ}$ is generated by a combination of the $\U{Y}$ and $\U{Z}$ generators. That means that all processes violating $\U{X} \times \U{YZ}$ (and in particular $\U{X}$) would be suppressed by some power of $\xi$. As we will see, in the limit that $\xi \ll 1$ it is possible to derive simple expressions for the masses and mixing matrices in the scalar sector. 
It is worth mentioning at this stage that, while such expressions serve as tools to understand the model's features, all scalar masses and mixing matrices 
will be computed fully numerically (i.e.~not as expansions in $\xi$) when scanning the parameter space of the model.

A spontaneously broken $\U{X} \times \U{Z}$ global symmetry would lead to massless Goldstone bosons and constrain the model significantly 
when considering e.g.~the precise measurements of the $Z$-boson width. This motivates us to softly break the symmetry by adding additional mass terms 
in the scalar potential. The scalar potential consistent with a softly broken $\U{X} \times \U{Z}$ global symmetry group can be split into fully symmetric 
and soft-breaking parts as $V=V_0 + V_{\mathrm{soft}}$, where
\begin{equation} 
\label{eq:V}
V_0 = - \sum_{i=1}^3 \mu_i^2 |H_i|^2 + \sum_{i,j=1}^{3} \left( \frac{\lambda_{ij}}{2} |H_i|^2 |H_j|^2 + \frac{\lambda'_{ij}}{2} |H_i^\dagger H_j|^2 \right) \, , \quad 
V_{\mathrm{soft}}=\sum_{i=1} ^3 \frac{1}{2}  ( m_{ij}^2 H^\dagger_i H_j + \mathrm{c.c} )
\end{equation}
with
\begin{align}
\lambda_{ij}=\lambda_{ji} \, , \quad \lambda'_{ij} = \lambda'_{ji}  \,, \quad m_{ij}^2 = m_{ji}^2 \, , \\
\lambda'_{11}=\lambda'_{22}=\lambda'_{33}=0 \, , \quad m_{11}^2 =m_{22}^2=m_{33}^2=0 \, .
\end{align}
All parameters in the scalar potential can be taken real without any loss of generality. This is due to the fact that the parameters in $V_0$ are real by construction, while any phases on $m_{ij}^2$ can be eliminated by field redefinitions of the three Higgs doublets. As a consequence the scalar sector of the model has no choice but to be CP-conserving. 

For convenience we define 
\begin{equation}
\tilde{\lambda}_{ij} =  (\lambda_{ij} + \lambda'_{ij}) \, .
\end{equation}

Finally, assuming that $v_{1,2,3} \neq 0$ and requiring that the first derivative of $V$ vanishes, we can write
\begin{equation}\label{eq:minimizationconds}
\mu_i^2 =\sum_{j=1}^3 \left[ \frac{1}{2}\tilde{\lambda}_{ij}  v_j^2 + m_{ij}^2\frac{v_j}{v_i} \right] \, .
\end{equation}

\subsection{Extending the $\U{X} \times \U{Z}$ to the fermion sector}

We assign the quark $\U{X} \times \U{Z}$ charges such that the neutral component of $H_3$ couples to only up- and down-type 
quarks of the third generation while the neutral components of $H_1$ and $H_2$ couple to the first and second generation down-type 
and up-type quarks respectively, i.e.
\begin{align}
\label{eq:LYukInt}
\mathcal{L}_{\mathrm{Yukawa}}^{\mathrm{q}} = \sum_{i,j=1}^2 \left\lbrace y^{\mathrm{d}}_{ij} 
\bar{d}_{\mathrm{R}}^i H_1^{\dagger} Q_{\mathrm{L}}^j - 
y^{\mathrm{u}}_{ij} \bar{u}_{\mathrm{R}}^i \tilde{H}_2^{\dagger} Q_{\mathrm{L}}^j \right\rbrace + 
y_{\mathrm{b}} \bar{b}_{\mathrm{R}} H_3^{\dagger} Q_{\mathrm{L}}^3 - 
y_{\mathrm{t}} \bar{t}_{\mathrm{R}} \tilde{H}^{\dagger}_3 Q_{\mathrm{L}}^3 + \mathrm{c.c.}  \,
\end{align}
In this way we forbid scalar-mediated tree-level FCNCs and simultaneously enforce a Cabibbo-like quark mixing, where the gauge eigenstates of the third quark family are aligned with the corresponding flavour eigenstates. This also means that a hierarchy in the VEVs of the Higgs doublets, where $v_3 \gg v_{1,2}$, leads to a third quark family that is much heavier than the first two without a strong hierarchy in the Yukawa couplings. 
\begin{table}
  \begin{center}
    \begin{tabular}{cccc}
     \hline \hline                     
                        				 	&$\mathrm{U}(1)_{\mathrm{Y}}$ & $\mathrm{U}(1)_{\mathrm{X}}$ & $\mathrm{U}(1)_{\mathrm{Z}}$  \\      
      \hline
       $H_1$								&$\frac{1}{2}$	&$-1$ 				&$-\frac{2}{3}$				\\
       $H_2$								&$\frac{1}{2}$	&$1$ 				&$\frac{1}{3}$				\\
       $H_3$								&$\frac{1}{2}$	&$0$ 				&$\frac{1}{3}$				\\ \hline
       $Q_{\mathrm{L}}^{1,2}$ 	&$\frac{1}{6}$	&$\gamma$		&$\delta$						\\
       $Q_{\mathrm{L}}^3$ 		&$\frac{1}{6}$	&$\beta$			&$\alpha$						\\ \hline
       $u_{\mathrm{R}}^{1,2}$ 	&$\frac{2}{3}$	&$1+\gamma$	&$\frac{1}{3}+\delta$		\\
       $t_{\mathrm{R}}$ 			&$\frac{2}{3}$	&$\beta$			&$\frac{1}{3}+\alpha$		\\	\hline
       $d_{\mathrm{R}}^{1,2}$ 	&$-\frac{1}{3}$	&$1+\gamma$	&$\frac{2}{3}+\delta$		\\
       $b_{\mathrm{R}}$ 			&$-\frac{1}{3}$	&$\beta$			&$-\frac{1}{3}+\alpha$		\\
     \hline \hline
    \end{tabular}
    \caption{Gharges of the global $\U{X}$, $\U{Z}$ and gauge (hypercharge) $\U{Y}$ symmetries in the considering class of 3HDMs. 
    The fermion charges together with the constraints in Eq.~\eqref{eq:ChargesConstraints} are chosen so that the only allowed Yukawa 
    terms are those in Eq.~\eqref{eq:LYukInt}.}
    \label{tab:U1Charges}
  \end{center}
\end{table}
In Table~\ref{tab:U1Charges}, we show the most general quark charge assignments allowing the terms in Eq.~\eqref{eq:LYukInt} once the $\U{X} \times \U{Z}$ 
charges of $H_{1,2,3}$ are fixed. As long as the parameters $\alpha$, $\beta$, $\gamma$ and $\delta$ in Table~\ref{tab:U1Charges} satisfy
\begin{equation} 
\label{eq:ChargesConstraints}
(\beta - \gamma ,  \alpha - \delta) \notin \lbrace (-1,-1),(-1,0),(0,0),(1,0),(1,1),(2,1) \rbrace 	\, , 
\end{equation}
the terms in Eq.~\eqref{eq:LYukInt} are also the \textit{only} allowed quark Yukawa interactions. It is worth noting that in the mass basis, the free parameters 
in the quark sector are simply the quark masses and the Cabibbo angle. The reader might note that at higher orders, the Yukawa interactions only allow for 
a mixing between the first and second quark generations, thus opening the question of how to reproduce the observed full CKM mixing in the quark sector. 
As this model is thought as an effective theory, one can write the following dimension-6 operators consistent with the imposed symmetries
\begin{equation}
\begin{aligned}
\bar{d}_{\mathrm{R}}^{1,2} \left(H_i^\dagger  Q_{\mathrm{L}}^3 \right) \left(H_j^\dagger H_k \right) \, , \quad \bar{u}_{\mathrm{R}}^{1,2} 
\left(\tilde{H}_i^\dagger  Q_{\mathrm{L}}^3 \right) \left(H_j^\dagger H_k \right) \, , \\
 \bar{b}_{\mathrm{R}} \left(H_i^\dagger  Q_{\mathrm{L}}^{1,2} \right) \left(H_j^\dagger H_k \right) \, , \quad \bar{t}_{\mathrm{R}} 
 \left(\tilde{H}_i^\dagger  Q_{\mathrm{L}}^{1,2} \right) \left(H_j^\dagger H_k \right) \, .
\end{aligned}
\end{equation}
Such terms will induce naturally small (suppressed by a scale of new physics) mixing terms with the third quark family once Higgs VEVs appear. 
The operators can in principle be generated \`a la Frogatt-Nielsen mechanism \cite{FROGGATT1979277} by integrating out the heavy fields of a high-energy theory. 
A deeper analysis of this is beyond the scope of this paper. 

Finally, we note that the lepton Yukawa sector can be made very SM-like by assigning the lepton $\U{X} \times \U{Z}$ charges such that they only 
couple to $H_3$. We will assume that this is the case throughout this work, and will not discuss the implications on lepton phenomenology any further. 
However, we want to point out that there are also other interesting scenarios, e.g.~where the leptons couple to $H_{1,2,3}$ such that the lepton mass 
hierarchies are also related to $v_{1,2} \ll v_3$.

\subsection{The spectrum, mixing matrices and interactions of the scalar sector}
\label{sec:SpecScalar}

After spontaneous symmetry breaking, the mass terms in the scalar potential $V$ in Eq.~\eqref{eq:V}, 
\begin{equation}
\sum_{i,j=1}^3 \left[\frac{1}{2} A_i (M_\mathrm{P}^2)_{ij} A_j  + \frac{1}{2} h_i (M_\mathrm{S}^2)_{ij} h_j + H^-_i (M_\mathrm{C}^2)_{ij} H^+_j  \right] \, , 
\end{equation}
can be neatly expressed using
\begin{equation}
\begin{aligned}
(M_\mathrm{P}^2)_{ij} &= m_{ij}^2- \delta_{ij} \sum_{k=1}^3 m_{ik}^2 \frac{v_k}{v_i} \, , 	\\
(M_\mathrm{S}^2)_{ij}  &= \tilde{\lambda}_{ij} v_i v_j + (M_{\mathrm{P}}^2)_{ij} \, ,		 \\
(M_\mathrm{C}^2)_{ij}  &= \lambda'_{ij} v_i v_j - 
\delta_{ij} \sum_{k=1}^3 \lambda'_{ik} v_k^2 + (M_{\mathrm{P}}^2)_{ij} \, .
\end{aligned}  
\label{eq:M2}
\end{equation}
We note that both $M_\mathrm{C,P}^2$ have an eigenvector $\propto v_i$ with vanishing eigenvalue. 
The corresponding Goldstone states become the longitudinal polarization states of the massive 
electroweak gauge bosons. 
 
The electrically neutral scalar, pseudo-scalar and charged scalar mass eigenstates,
\begin{equation}
\label{eq:MassStates}
\bar{h}_i = (h_{\mathrm{a}}, h_{\mathrm{b}}, h_{125}) \, , \quad 
\bar{A}_i = (A_{\mathrm{a}}, A_{\mathrm{b}}, A_{\mathrm{G}}) \, , \quad 
\bar{H}^{\pm}_i = (H^\pm_{\mathrm{a}} , H^\pm_{\mathrm{b}}, H^{\pm}_{\mathrm{G}}) \, , 
\end{equation} 
are related to the interaction eigenstates as
\begin{equation} 
\label{eq:ScalarMixingsDefinition}
h_i = \mathsf{S}_{ij} \bar{h}_j \, , \quad A_i = \mathsf{P}_{ij} \bar{A}_j \, , \quad 
H^\pm_i = \mathsf{C}_{ij} \bar{H}_j^\pm \, , 
\end{equation}

The states $A_{\mathrm{G}}$ and $H^{\pm}_{\mathrm{G}}$ in Eq.~\eqref{eq:MassStates} denote the Goldstone bosons. 
Working in the $\xi \ll 1$ limit, the mixing matrices $\mathsf{S}$, $\mathsf{P}$ and $\mathsf{C}$ are identical up to 
$\mathcal{O}(\xi)$ but differ at $\mathcal{O}(\xi^2)$. It is here convenient to define an angle $\beta\in [0,\frac{\pi}{2}]$ as
\begin{equation}
\tan \beta = \frac{v_2}{v_1} \, .
\end{equation}
To the second order in $\xi$, we have
\begin{equation}
\mathsf{S} = \mathsf{T} + \xi^2 \mathsf{S}' \, , \quad \mathsf{P}=\mathsf{T} + \xi^2 \mathsf{P}' \, , \quad 
\mathsf{C} = \mathsf{T} + \xi^2 \mathsf{C}'
\end{equation}
with
\begin{equation}
\mathsf{T} = \begin{psmallmatrix}
1 & X \xi & c_\beta \xi  		\\
-X \xi & 1 & s_\beta	 \xi 	\\
- c_\beta \xi  & - s_\beta \xi  & 1  
\end{psmallmatrix} \, , \quad X \equiv 
\frac{m_{12}^2 s_\beta c_\beta}{m_{13}^2 s_\beta - m_{23}^2 c_\beta} \, . 
\end{equation}
For the $\mathcal{O}(\xi^2)$ pieces, we have
\begin{equation}
\mathsf{P}' = \begin{psmallmatrix}
-\frac{1}{2} (X^2+c_\beta^2) & - \frac{1}{2}(1-Y)s_\beta c_\beta & 0 \\
-\frac{1}{2} (1+Y)s_\beta c_\beta & -\frac{1}{2}(X^2 + s_\beta^2) & 0 \\
X s_\beta & -X c_\beta & 0 
\end{psmallmatrix}  \, ,
\mathsf{C}' =\mathsf{P}' + \begin{psmallmatrix}
0 & Z_1 & 0 \\
-Z_1 & 0 & 0 \\
0 & 0 & 0  
\end{psmallmatrix} \, ,\mathsf{S}' = \mathsf{P}' + \begin{psmallmatrix}
0 & 0 & Z_2 \\
0 & 0 & Z_3 \\
-Z_2 & -Z_3 & 0
\end{psmallmatrix} 
\, , 
\end{equation}
where
\begin{equation}
\begin{aligned}
Y &= \frac{ (2 m_{12}^4 + m_{23}^4) c_\beta^2 - (2 m_{12}^4 + m_{13}^4) s_\beta^2}{(m_{13}^2 s_\beta - m_{23}^2 c_b^2)^2}	\, &, \quad  
Z_1 &= \frac{(\lambda_{23}'-\lambda_{13}')s_\beta^2 c_\beta^2 m_{12}^2 v_{3}^2}{(m_{13}^2 s_\beta - m_{23}^2 c_{\beta})^2}	\, , 	\\
Z_2 &= (\tilde{\lambda}_{13}-\lambda_{33}) c_\beta^2 \frac{v_3^2}{m_{13}^2} \, &, 
\quad 	Z_3 &= (\tilde{\lambda}_{23}-\lambda_{33}) s_\beta^2 \frac{v_3^2}{m_{23}^2}  \, . 	
\end{aligned}
\end{equation}
Here, $Z_{1,2,3}$ parametrize the leading order difference between the mixing matrices, which will be important as these parameters determine 
the off-diagonal scalar-scalar interactions with the electroweak gauge bosons. We also note that as $X$, $Y$, $Z_1$, $Z_2$, $Z_3$ get larger, 
the expansion in $\xi$ becomes less reliable. 

The state $h_{125}$ contains mostly $h_3$, meaning that it couples substantially to the third quark family. 
It also receives a mass of the order of $v_3 \sim v$,
\begin{equation}
m_{h_{125}}^2 =  \lambda_{33} v_3^2 + \mathcal{O} (\xi^2 ) \, ,
\end{equation}
making this state our candidate for the observed SM Higgs-like 125 GeV state. The exotic scalars $h_{\mathrm{a,b}}$, $A_{\mathrm{a,b}}$ and 
$H^\pm_{\mathrm{a,b}}$ can all be made heavy as the leading order contribution to their masses is inversely proportional to $\xi$. 
To this leading order, $\lbrace h_{\mathrm{a}}, A_{\mathrm{a}}, H^\pm_{\mathrm{a}} \rbrace$ are degenerate in mass. 
This is also the case for $\lbrace h_{\mathrm{b}}, A_{\mathrm{b}}, H^\pm_{\mathrm{b}} \rbrace$.  
More accurately, the masses are given by
\begin{equation} 
\label{eq:ScalarMasses}
\begin{aligned}
m_{A_{\mathrm{a}}}^2 = m_{h_{\mathrm{a}}}^2 = - \frac{m_{13}^2}{c_\beta \xi}-m_{12}^2 t_\beta - (m_{13}^2 c_\beta +  X m_{12}^2) \xi \, &, \quad 
m_{H_{\mathrm{a}}^\pm}^2 =	m_{A_{\mathrm{a}}}^2-\lambda_{13}' v_3^2	\, ,			\\
m_{A_{\mathrm{b}}}^2 = m_{h_{\mathrm{b}}}^2 = - \frac{m_{23}^2}{s_\beta \xi}-\frac{m_{12}^2}{t_\beta} - (m_{23}^2 s_\beta +  X m_{12}^2) \xi \, &, \quad 
m_{H_{\mathrm{b}}^\pm}^2 =	m_{A_{\mathrm{b}}}^2-\lambda_{23}' v_3^2 \, , 
\end{aligned}
\end{equation}
to $\mathcal{O}(\xi)$. This means that the exotic scalars and pseudo-scalars are typically very close in mass when $\xi \ll 1$, i.e.~$m_{A_{\rm{a,b}}}^2 - m_{h_{\rm{a,b}}}^2 = \mathcal{O}(\xi^2)$. Note also that the couplings $\lambda_{ij}'$ can either be positive or negative, such that the charged scalars $H^+_{\rm{a,b}}$ can both be heavier and lighter than the neutral scalars in the respective family.  

We conclude this section by listing the trilinear interactions between the physical scalars and the electroweak gauge bosons, as they are relevant 
for the collider phenomenology of the charged Higgs boson discussed in section~\ref{sec:hcprod}. The interactions between the neutral scalars, charged scalars and the $W$ boson 
are given by
\begin{equation} 
\label{eq:lagWH}
\begin{aligned}
\mathcal{L} &\supset  \mathrm{i} \frac{g_2}{2} W^-_{\mu}  \sum_{i=1}^3\left[ (\partial^{\mu} H_i^+ ) h_i - H^+_i (\partial^{\mu} h_i) \right] + \mathrm{c.c} \\
&=  \mathrm{i} \frac{g_2}{2} W^-_{\mu}  \sum_{i,j=1}^3 (\mathsf{C}^{\mathrm{T}} \mathsf{S})_{ij} \left[ (\partial^{\mu} \bar{H}_i^+) \bar{h}_j - 
\bar{H}^+_i (\partial^{\mu} \bar{h}_j) \right] + \mathrm{c.c} \, ,
\end{aligned} 
\end{equation}
with
\begin{equation} 
\label{eq:WHhCoupling}
\mathsf{C}^{\mathrm{T}} \mathsf{S} = \begin{psmallmatrix}
1 				& - Z_1\xi^2 	&  Z_2 \xi^2	\\
 Z_1 	\xi^2	& 1 				& Z_3 \xi^2	\\
- Z_2 \xi^2	& - Z_3 \xi^2	& 1
\end{psmallmatrix} + \mathcal{O}(\xi^3) \, . 
\end{equation}
The top line in Eq.~\eqref{eq:lagWH} is written in the interaction eigenbasis of the scalars, while the bottom line is the same expression in terms of the mass eigenstates. The $W$ boson also couples to pairs of charged scalars and pseudo-scalars as 
\begin{equation} 
\label{eq:lagWA}
\begin{aligned}
\mathcal{L} &\supset \frac{g_2}{2} W^-_{\mu}  \sum_{i=1}^3 \left[ (\partial^{\mu} H_i^+ ) A_i - H^+_i (\partial^{\mu} A_i) \right] + \mathrm{c.c} \\
&= \frac{g_2}{2} W^-_{\mu} \sum_{i,j=1}^3 (\mathsf{C}^{\mathrm{T}} \mathsf{P})_{ij} \left[ (\partial^{\mu} \bar{H}_i^+) 
\bar{A}_j - \bar{H}^+_i (\partial^{\mu} \bar{A}_j) \right]  + \mathrm{c.c} \, ,
\end{aligned} 
\end{equation}
with
\begin{equation} 
\label{eq:WHACoupling}
\mathsf{C}^{\mathrm{T}} \mathsf{P} = \begin{psmallmatrix}
1 				& - Z_1\xi^2 	& 0	\\
 Z_1 	\xi^2	& 1 				& 0	\\
0	& 0	& 1
\end{psmallmatrix} + \mathcal{O}(\xi^3) \, . 
\end{equation}
Similarly, for the trilinear interactions with the $Z$ boson, we have
\begin{equation}
\begin{aligned}
\mathcal{L} &\supset \frac{g_2}{2 c_\mathrm{W}} Z_{\mu} \sum_{i=1}^3\left[ (\partial^{\mu} A_i ) h_i - 
A_i (\partial^{\mu} h_i) \right]  \\
&=  \frac{g_2}{2 c_\mathrm{W}}  Z_{\mu}  \sum_{i,j=1}^3 (\mathsf{P}^{\mathrm{T}} \mathsf{S})_{ij} \left[ (\partial^{\mu} \bar{A}_i) \bar{h}_j - 
\bar{A}_i (\partial^{\mu} \bar{h}_j) \right],
\end{aligned}
\end{equation}
where $c_{\mathrm{W}}$ is the cosine of the Weinberg angle and
\begin{equation} \label{eq:ZAhCoupling}
\mathsf{P}^{\mathrm{T}} \mathsf{S} = \begin{psmallmatrix}
1 				& 0 	& Z_2 \xi^2 	\\
0 	& 1 				& Z_3 \xi^2	\\
- Z_2 \xi^2	& - Z_3 \xi^2	& 1
\end{psmallmatrix} + \mathcal{O}(\xi^3) \, . 
\end{equation}

\subsection{Scalar-fermion couplings}

Knowing the mixing matrices $\mathsf{S}$, $\mathsf{P}$ and $\mathsf{C}$ for the neutral scalars, pseudo-scalars and charged scalars, respectively,
to the first orders in $\xi$, it is straightforward to obtain the Yukawa interactions between the physical scalars and the quarks. 
Using Eq.~\eqref{eq:ScalarMixingsDefinition} we find that $h_{125}$ couples to quarks in a way similar to the SM, 
\begin{equation}\label{eq:SMHiggsY}
\mathcal{L} \supset \sum_{q} \frac{m_q}{v_3} \, \bar{q} q \, h_{125} + \mathcal{O}(\xi) \, .
\end{equation}
For the third quark family, this is an obvious consequence of the model's symmetries, as $t$ and $b$ quarks receive their masses from $H_3$ with 
$v_3 \lesssim v$, and $h_{125}$ is mostly made of $h_3$. On the other hand, the first and second family get their masses from $H_{1,2}$ with $v_{1,2} \ll v_3 $, 
so the corresponding Yukawa couplings with the gauge eigenstates $H_{1,2}$ are quite large as $\mathcal{O}(m_q/v_{1,2}) \sim \mathcal{O}(m_q / \xi v_3)$. 
When shifting to the mass eigenbasis, $h_{1,2}$ contribute to $h_{125}$ only at $\mathcal{O}(\xi)$ thus giving an overall coupling of $\mathcal{O}(m_q/v_3)$.

In the same process, we also find the interaction terms between the quarks and the exotic scalar states $h_{\mathrm{a,b}}$, $A_{\mathrm{a,b}}$ and 
$H_{\mathrm{a,b}}^\pm$. Couplings to the third quark family are generally quite small $\sim m_{\mathrm{t,b}}\xi/v_3$. In our model,
phenomenologically the most relevant couplings are with the second quark family instead, which to the leading order in $\xi$ read
\begin{equation}\label{eq:ExHiggsY}
\begin{aligned}
\mathcal{L} \supset &  \cos{\theta_{\mathrm{C}}} \frac{\sqrt{2} m_{\mathrm{s}}}{v_1} \bar{s}_{\mathrm{R}} c_{\mathrm{L}} H^-_{\mathrm{a}} - 
 \cos{\theta_{\mathrm{C}}} \frac{\sqrt{2} m_{\mathrm{c}}}{v_2} \bar{c}_{\mathrm{R}} s_{\mathrm{L}} H^+_{\mathrm{b}} + \mathrm{c.c.}	 + \mathcal{O}(\xi)		\\
&  + \frac{m_\mathrm{s}}{v_1} \bar{s} s h_{\mathrm{a}} - \frac{m_\mathrm{c}}{v_2} \bar{c} c h_{\mathrm{b}} + 
\mathrm{i} \frac{m_\mathrm{s}}{v_1} \bar{s} \gamma^5 s A_\mathrm{a} - \mathrm{i}  \frac{m_\mathrm{c}}{v_2} \bar{c} \gamma^5 c A_\mathrm{b} + \mathcal{O}(\xi) \, ,
\end{aligned}
\end{equation}
where $\theta_{\mathrm{C}}$ is the Cabbibo angle. When the masses of the scalars are in the appropriate range, we can expect that the charged scalars 
$H^+_{\mathrm{a,b}}$ would be produced in collider experiments through $c \bar{s}$ fusion while $h_\mathrm{a}$ and $A_\mathrm{a}$ ($h_\mathrm{b}$ and 
$A_\mathrm{b}$) would mainly be produced by the $s \bar{s}$ ($c \bar{c}$) fusion.

\section{A model independent approach} 
\label{sec:hcprod}

One of the interesting features of our model is the existence of heavy charged scalars $H^+$ ($H^-$) that mostly couple to a $c \bar s$ ($\bar cs$) pair 
as their interactions with $t \bar b$ ($\bar t b$) are small due to the model symmetries. Furthermore, we find that $H^\pm$ can decay to a $W^\pm\,h_{125}$ pair 
with a sizable branching ratio (BR) which is still allowed by the current experimental data. It turns out that this unconventional channel, while not explored 
in the literature before for $m_{H^\pm} > 200$ GeV, can be a rather clean way to search for charged scalars at the LHC. 

In the following, we adopt a model independent approach in searching for charged scalars exhibiting those features. In section \ref{sec:translation}, we will 
show how the analysis can be used to find discovery regions in the parameter space of the 3HDM we have proposed above. We take a model independent 
approach to not only test the predictions of our model, but also to offer a guideline for our experimental colleagues to implement this new search channel in 
the experimental analyses. 

\begin{figure}[!htbp]
\includegraphics[height=6cm]{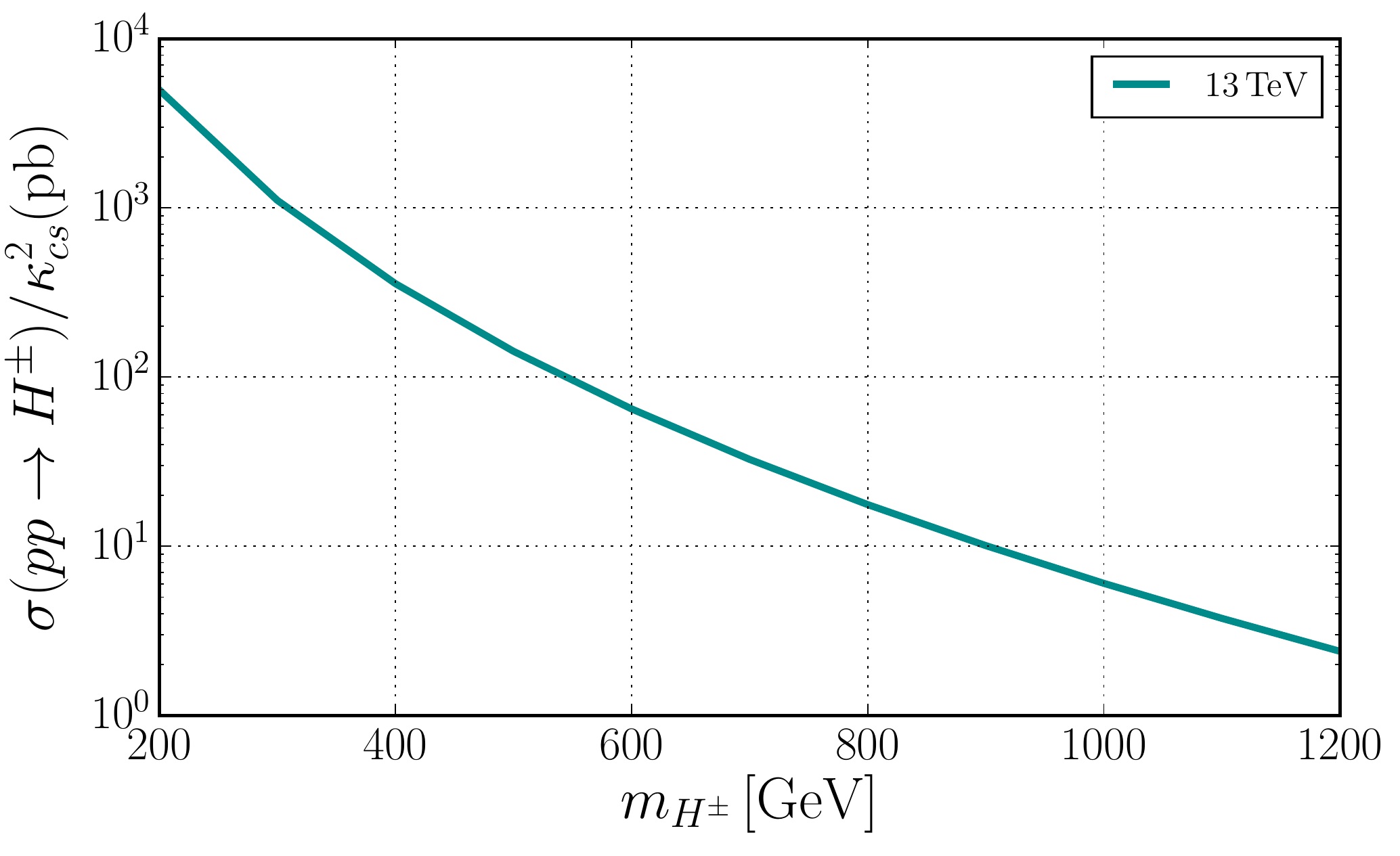}
\caption{$\sg(pp\to H^\pm) / \kp_{cs}^2 = \sg_0(m_{H^\pm})$ as a function of $m_{H^\pm}$ at the LHC ($\sqrt{s}=13$ TeV). }
\label{fig:prodcs}
\end{figure}

We start with the following model independent Lagrangian for $H^\pm$ including its kinetic $(\mc{L}_{kin})$ and interaction $(\mc{L}_{int})$ terms
\begin{align} 
\label{eq:lagkin}
\mc{L}_{kin} \supset D_{\mu}H^{+}D^{\mu}H^{-} - m_{H^\pm}^2~H^{+}H^{-} \,,
\end{align}
\begin{align}
\label{eq:lagint}
\mc{L}_{int} \supset \kp_{cs}^p~\bar{c}_{\rm R}s_{\rm L}H^{+}
+ \kp_{cs}^m~\bar{s}_{\rm R}c_{\rm L}H^{-} + i\kp_{Wh_{125}}~\lt(h_{125}\partial^{\mu}H^{+} - 
H^{+}\partial^{\mu}h_{125}\rt) W^-_{\mu} + \textrm{c.c.} \,.
\end{align}
There are four free parameters in the above Lagrangian \emph{viz}.~the charged Higgs mass $m_{H^\pm}$, and the three 
couplings $\kp_{cs}^p$, $\kp_{cs}^m$ and $\kp_{Wh_{125}}$.
In general, $\kp_{cs}^p$ and $\kp_{cs}^m$ both could be non-zero. In that case, the production cross section, $\sg(pp\to H^\pm)$ 
is proportional to the combination $\lt[\lt(\kp_{cs}^p\rt)^2 + \lt(\kp_{cs}^m\rt)^2\rt]$. Therefore, instead of two free couplings, we introduce 
a single free parameter $\kp_{cs}$ which is, $\kp_{cs}^2 = \lt(\kp_{cs}^p\rt)^2 + \lt(\kp_{cs}^m\rt)^2$. From the above model independent 
Lagrangian, we see that $H^+$ has only two decay modes: $W^+\,h_{125}$ and $c\bar s$. The corresponding tree-level partial widths are given by
\begin{align}
\Gm\lt(H^\pm\to W^\pm\, h_{125}\rt) &= \frac{\kp_{Wh_{125}}^2m_{H^\pm}^3}{64\pi m_W^2}\lt[1-\frac{\lt(m_{h_{125}}-m_W\rt)^2}{m_{H^\pm}^2}\rt]
\lt[1-\frac{\lt(m_{h_{125}}+m_W\rt)^2}{m_{H^\pm}^2}\rt]\nn\\
&\times\lt[1-\frac{2\lt(m_{h_{125}}^2+m_W^2\rt)}{m_{H^\pm}^2} + \frac{\lt(m_{h_{125}}^2-m_W^2\rt)^2}{m_{H^\pm}^4}\rt]^{1/2}\,,\\
\Gm\lt(H^+\to c\bar s\rt) &= \frac{3\lt[\lt(\kp_{cs}^p\rt)^2 + \lt(\kp_{cs}^m\rt)^2\rt]m_{H^\pm}}{16\pi} 
= \frac{3\kp_{cs}^2 m_{H^\pm}}{16\pi} \,.
\end{align}
where $m_{h_{125}}=125$ GeV. The expression $\Gamma(H^+\to c\bar{s})$ is given in the limit of massless $c$ and $s$ quarks. In general, 
$H^\pm$ can have other decay modes too. We, therefore, take the BR of the decay mode $H^\pm\to W^\pm \,h_{125}$ denoted by 
${\rm BR}_{Wh_{125}}$ as a free parameter instead of $\kp_{Wh_{125}}$. So, one can write the following in the narrow-width approximation,
\begin{align}
\label{eq:hcprodsig}
\sg(pp\to H^\pm \to W^\pm\,h_{125}) = \sg(pp\to H^\pm)\times {\rm BR}_{Wh_{125}} 
= \kp_{cs}^2\times \sg_0(m_{H^\pm})\times {\rm BR}_{Wh_{125}} \,,
\end{align} 
where $\sg_0(m_{H^\pm})$ is the cross section of $pp\to H^\pm$ for $\kp_{cs}=1$. We show 
$\sg_0(m_{H^\pm})$ at the LHC ($\sqrt{s}=13$ TeV) as a function of $m_{H^\pm}$ in Fig.~\ref{fig:prodcs}.

\section{Search for charged scalars produced by $c \bar{s}$ fusion}
\label{sec:ColliderPheno}

We implement the model independent Lagrangian of $H^\pm$ as shown in Eqs.~\eqref{eq:lagkin} and  
\eqref{eq:lagint} in \textsc{FeynRules}~\cite{Alloul:2013bka} from which we get the Universal FeynRules 
Output~\cite{Degrande:2011ua} model files for the \textsc{MadGraph}~\cite{Alwall:2014hca} event generator. 
We use the NNPDF~\cite{Ball:2012cx} parton distribution functions (PDFs) for the signal and background event 
generation. For the signal, we use fixed factorization $\mu_F$ and renormalization $\mu_R$ scales at 
$\mu_F=\mu_R=m_{H^\pm}$ while for the background these scales are chosen at the appropriate scale of the process. 
We use \textsc{Pythia6}~\cite{Sjostrand:2006za} for subsequent showering and hadronization of the generated events. 
Detector simulation is performed using \textsc{Delphes} \cite{deFavereau:2013fsa} which employs the \textsc{FastJet} 
\cite{Cacciari:2011ma} package for jet clustering. Jets are clustered using the anti-kT algorithm \cite{Cacciari:2008gp} 
with the clustering parameter $R = 0.4$. For the multivariate analysis (MVA), we use the Boosted Decision Tree (BDT) 
algorithm in the TMVA \cite{Hocker:2007ht} framework. In this analysis, all calculations are done at the leading order, 
for simplicity.

\subsection{Signal}
\label{subsec:sig}

We focus the $H^+$ ($H^-$) production from the $c\bar s$ ($\bar c s$) initial state followed by the decay 
$H^\pm\to W^\pm\,h_{125}$. We consider a semileptonic final state where $W^\pm$ decays leptonically and $h_{125}$
decays to $b\bar{b}$. Therefore, the chain of the signal process in our case is 
\begin{align}
\label{eq:sigpross}
pp\to H^{\pm} \to W^{\pm}h_{125}\to \ell^{\pm}+\slashed{E}_T + b\bar{b}\ .
\end{align}
Here, $\ell=\{e,\mu\}$. We then have one charged lepton, two $b$-jets and missing transverse energy in the final 
state and our event selection criteria is exactly one charged lepton (either an electron or a muon including their 
anti-particles), at least two jets and missing transverse energy that pass the following basic selection cuts:
\begin{itemize}
\item Lepton: $p_{T}(\ell) > 25$ GeV, $|\eta(\ell)| < 2.5$
\item Jet: $p_{T}(J) > 25$ GeV, $|\eta(J)| < 4.5$
\item Missing transverse energy: $\slashed{E}_T > 25$ GeV
\item $\Dl R$ separation: $\Dl R(J_1,J_2) > 0.4$, $\Dl R(\ell,J) > 0.4$
\end{itemize}
Here, $J_1$ and $J_2$ denote the first and the second highest $p_T$ jets. After selecting the events, we further demand 
$b$-tagging on the two leading-$p_T$ jets. The $b$-tagging on jets can reduce the background very effectively but 
it can also somewhat reduce the signal. Therefore, to enhance the signal cut efficiency we do not always demand two $b$'s tagging 
although there are two $b$-jets present in the signal. Depending on the number of $b$-tagged jets we demand, we define 
the following two signal categories
\begin{itemize}
\item \underline{$1b$-tag:} In this category, we demand at least one $b$-tagged jet among the two leading $p_T$ jets.
\item \underline{$2b$-tag:} In this category, we demand that both the two leading $p_T$ jets are $b$-tagged. 
This category is a subset of the $1b$-tag category.
\end{itemize}

To reconstruct the Higgs boson, we apply an invariant mass cut $|m_{H^\pm}-m_{h_{125}}|< 20$ GeV around 
the Higgs boson mass $m_{h_{125}}=125$ GeV. However, the full event is not totally reconstructible due to the presence 
of the missing transverse energy. 

\subsection{Background}
\label{subsec:back}

The main background for the signal with one lepton, at least one or two $b$-tagged jets and missing energy 
can come from the following SM processes:
\begin{table}
\centering
\begin{tabular}{ccccccccccc}
\hline \hline
Process & $\,\, W+n~j \,\,$ & $\,\, Wbj \,\, $ & $\,\, Wb\bar{b} \,\,$ & $\,\, t\bar{t}+n~j \,\,$ & $\,\, tj \,\,$ 
& $\,\,\,\, tb \,\,\,\,$ & $\,\, tW \,\,$ & $\,\, WW \,\,$ & $\,\, WZ \,\,$ & $\,\, Wh_{125}\,\,$ \\ 
\hline 
x-sec (pb) & $1.53\times 10^5$ & 308.9 & 41.7 & 431.3 & 174.6 & 2.6 & 54.0 & 67.8 & 25.4 & 1.1 \\ 
\hline \hline
\end{tabular} 
\caption{Parton-level cross sections of various background processes (without any cut) at the LHC ($\sqrt{s}=13$ TeV). Here, $n$ is the number of jets.}
\label{tab:Bcs}
\end{table}
\begin{enumerate}

\item \underline{$W^\pm+\textrm{jets}$}: 
The definition of our inclusive $W^\pm+\textrm{jets}$ background includes up to two jets and we include the
$b$ parton in the jet definition i.e.~$j=\{g,u,d,c,s,b\}$. We generate these background events in two 
separate parts. In one sample, we only consider light jets i.e.$j=\{g,u,d,c,s\}$ and combine $pp\to W^\pm+(0,1,2)~j$ 
processes where we set the matching scale $Q_{cut}=25$ GeV. This background is the largest
(the cross section is about $1.53\times 10^5$ pb at the LHC, with $\sqrt{s}=13$ TeV, without any cut)
among all the dominant SM backgrounds we have considered. Although the bare cross section is large,
it will reduce drastically after $b$-tagging due to a small mistagging (light jet is tagged as $b$-jet) rate. 
We find that its contribution in the $1b$-tag category is substantial but in the $2b$-tag category is very small.
In the other sample, we consider at least one $b$ parton in the final state where we combine $pp\to W^\pm\,bj$
and $pp\to W^\pm\, b\bar b$ processes (no SM $pp\to W^\pm\,b$ process exists). This background will 
contribute significantly in both the categories. We include $pp\to W^\pm\,h_{125}\to W^\pm\,b\bar b$ 
and $pp\to W^\pm\,Z\to W^\pm\, b\bar b$ processes in the $pp\to W^\pm\,b\bar b$ channel.

\item \underline{$t\bar{t}+\textrm{jets}$:} 
The definition of our inclusive $t\bar t+\textrm{jets}$ background includes up to two jets containing also $b$ partons. 
We generate this background by combining $pp\to t\bar t+(0,1,2)~j$ processes using the matching scale $Q_{cut}=25$ GeV. 
The matched cross section is about 431 pb before the top decay and without any selection cut applied. We find that this background is 
the dominant one after the strong basic selection cuts (applied before passing the events to the MVA).

\item \underline{Single top:} 
This background includes three types of single top processes -- $s$-channel single top (such $pp\to t\bar b$), $t$-channel 
single top (i.e.~$pp\to tj$) and single-top associated with $W$ (such as $pp\to tW^\pm$) processes. Note that for the $pp\to tW^\pm$ process, 
the selected lepton can come from two possible ways, either from the decay of the associated $W^\pm$ or from the $W^\pm$ coming from 
the top decay. These two possibilities are properly included in our event sample. The single top background also contributes significantly 
to the total background. 

\item \underline{Diboson:}
This background includes $pp\to W^\pm\, W^\mp\to W^\pm+jj$ and $pp\to W^\pm\,Z\to W^\pm+jj$ processes where two light jets 
come from the decay of $W$ or $Z$ bosons. In this background, we have also included $pp\to W^\pm\,Z\to W^\pm\,\nu\bar{\nu}$ processes 
where two selected jets come from the parton showers. This background reduces drastically due to the small mistagging efficiency of light jets that are 
misidentified as $b$-jets. Finally, in the MVA this background contributes negligibly to the total background. Note that two diboson production processes \emph{viz}.~$pp\to W^\pm h_{125}\to W^\pm b\bar{b}$ and $pp\to W^\pm Z\to W^\pm b\bar{b}$ are already considered in the $W+\textrm{jets}$
background.  

\item \underline{QCD multijets:}
The multijet background arises due to QCD interactions at the LHC and has a very large production cross section, especially in the soft region. 
The QCD-induced multijet production processes can potentially contribute to the total background for our signal by faking the lepton, 
$\slashed{E}_T$ and $b$-tagged jets. It is impractical to study this part of the background using a Monte-Carlo simulation since it is computationally 
challenging to generate enough events due to very low fake rates. In experimental analyses, this contribution is usually estimated from the data. 
In our analysis, we do not consider this background since it will be largely diminished after strong preselection cuts and will be further reduced 
due to small fake rates of the considered final states.

\end{enumerate}

The SM background, especially the $W+\textrm{jets}$ component, is large and therefore one has to design a clever set of cuts which 
would notably reduce such a background but would not notably affect the signal. This implies that the cut efficiency for the background is very small and hence, 
a large number of background events has to be generated. In order to avoid the generation of a large event sample, we apply a strong cut on 
the partonic center-of-mass energy, $\sqrt{\hat{s}} > 200$ GeV at the generation level of all background processes. This cut can reduce the $W+\textrm{jets}$ 
background by two orders of magnitude. However, this cut has no or very little effect on the other backgrounds \emph{viz}.~$t\bar{t}+\textrm{jets}$, single top 
and diboson ones since the threshold energy for them is either above or slightly below 200 GeV. In the case of a signal, $\sqrt{\hat{s}}$ is always above 
200 GeV since we are interested in the parameter space regions where $m_{H^\pm}> m_W+m_{h_{125}}\gtrsim 205$ GeV. 

One should note that, in reality, the full reconstruction of $\sqrt{\hat{s}}$ of an event is not possible if there is missing energy present in that event. 
In this case, one can construct an inclusive global variable $\sqrt{\hat{s}_{min}}$ defined in Ref.~\cite{Konar:2008ei} which is closest to the actual 
$\sqrt{\hat{s}}$ of the event. One can roughly approximate $\sqrt{\hat{s}}\approx \sqrt{\hat{s}_{min}}$ if there is only one missing neutrino in the event 
but this approximation gets poorer with the increase of the number of neutrinos in the final state. For simplicity, we have used the cut $\sqrt{\hat{s}} > 200$ GeV 
at the generation level. But in reality, one can use a cut on $\sqrt{\hat{s}_{min}}$ to trim the background before passing it for further analysis.

\subsection{Multivariate analysis}

A $Wh_{125}$ resonance, similar to our case, can also appear from the decay of a heavy charged gauge boson, $W'$. The search for $W'$ in the 
$\ell^{\pm}+\slashed{E}_T + b\bar{b}$ channel (same final state that we are interested in) has been carried out at the LHC~\cite{ATLAS:2017nxi,Sirunyan:2017nrt}. 
In these searches, they mainly focus in the TeV-scale $W'$ mass and the analyses are done using the cut-based techniques. A cut-based analysis may not perform 
well in our case, especially for low $m_{H^\pm}$ region due to the presence of a large SM background~\cite{Li:2016umm,Patrick:2016rtw}. Therefore, we choose 
to use a MVA to obtain a better signal-to-background discrimination which usually leads to a better significance than a cut-based analysis. See Ref.~\cite{Bhat:2010zz} 
for a brief review on various multivariate methods and their use in collider searches. In this paper, we only use multivariate techniques and do not compare our 
achieved sensitivity with the cut-based techniques. 

\begin{figure}[!htbp]
\includegraphics[height=9cm,width=16cm]{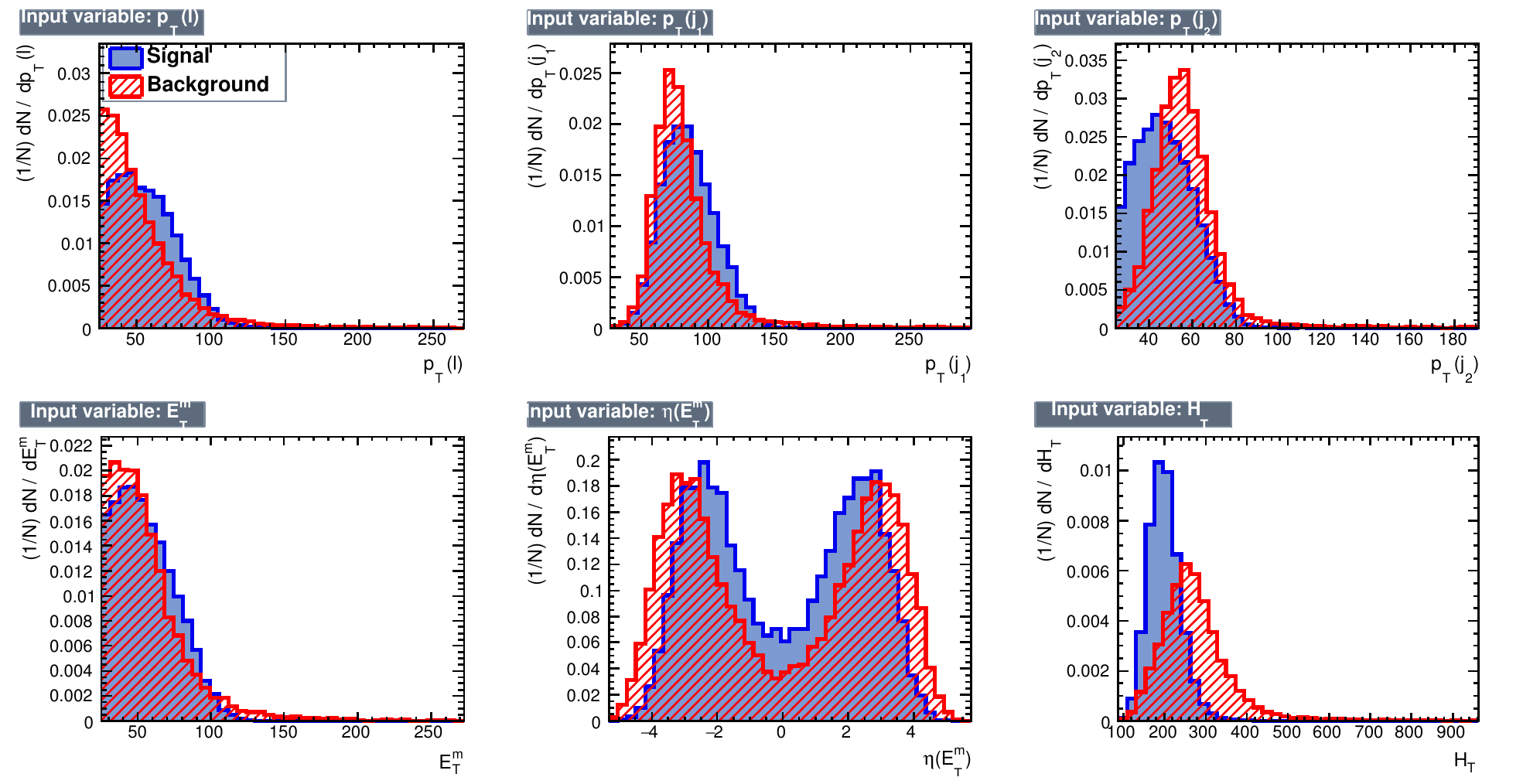}\\
\includegraphics[height=9cm,width=16cm]{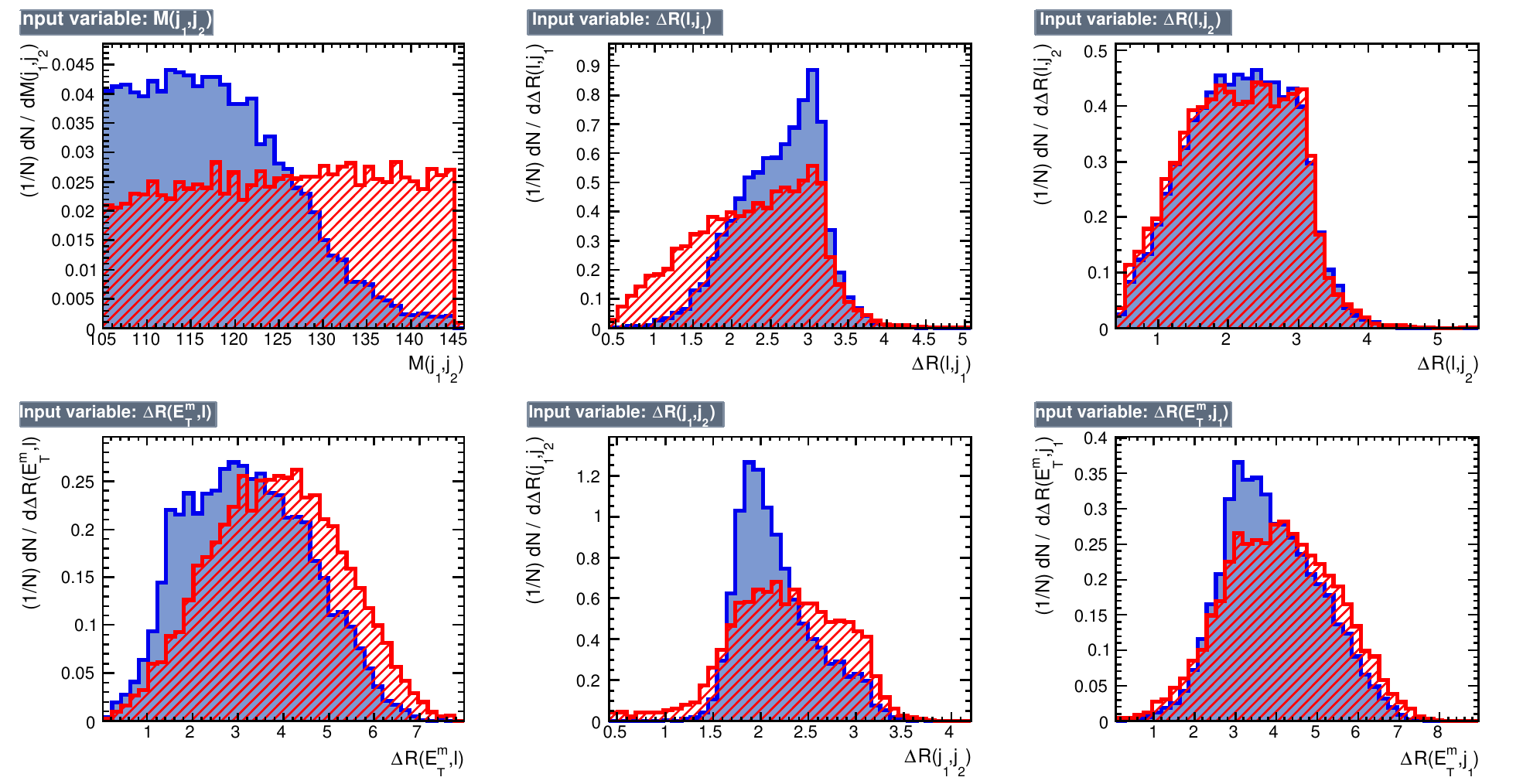}
\caption{Normalized distributions of the input variables at the LHC ($\sqrt{s}=13$ TeV) used in the MVA for the signal (blue) and the background (red). 
Signal distributions are obtained for $m_{H^\pm}=300$ GeV, and the background includes all the dominant backgrounds discussed in 
subsection~\ref{subsec:back}. These distributions are drawn by selecting events after the cuts defined in subsection
\ref{subsec:sig}.}
\label{fig:var300}
\end{figure}

\begin{figure}[!htbp]
\includegraphics[height=9cm,width=16cm]{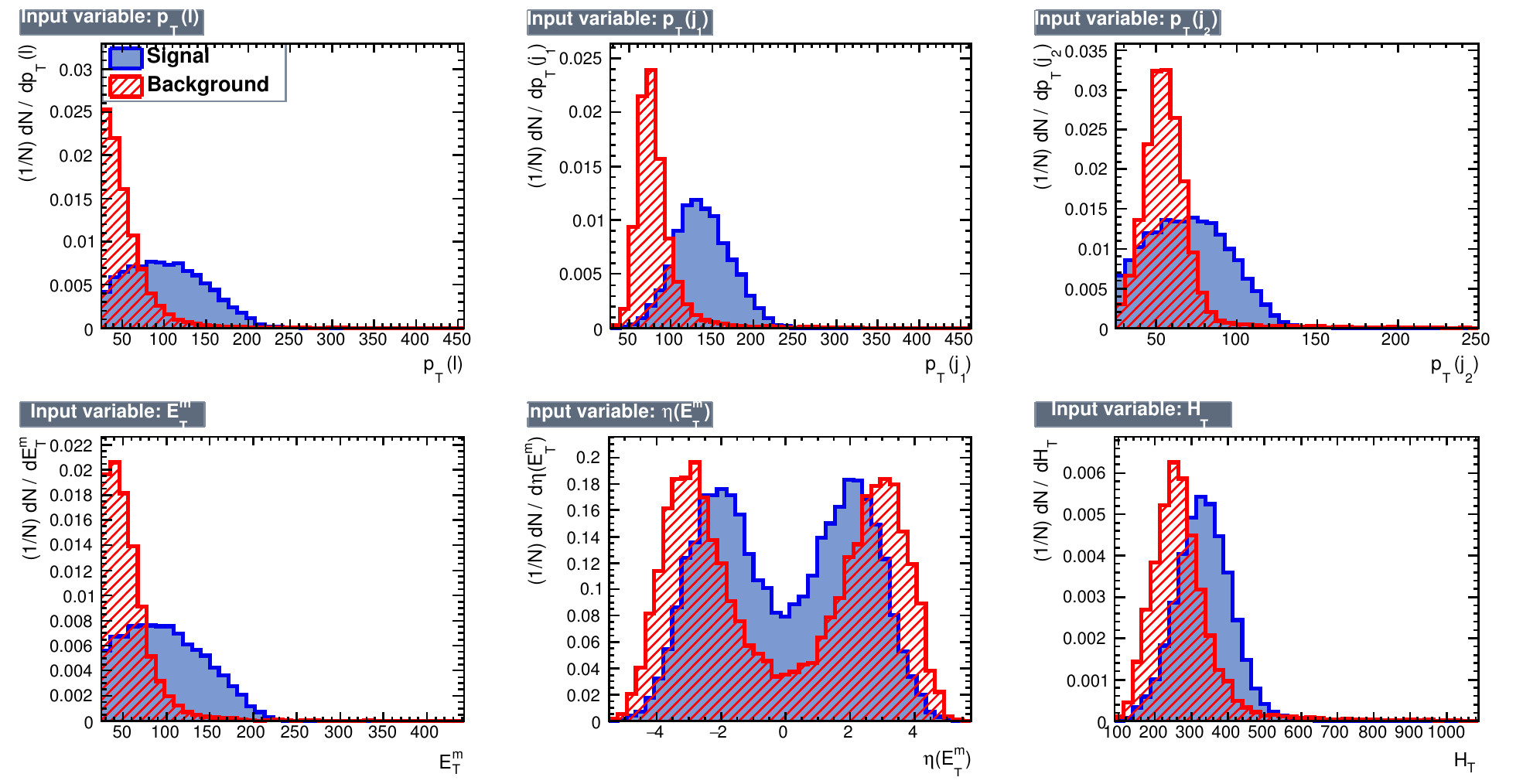}\\
\includegraphics[height=9cm,width=16cm]{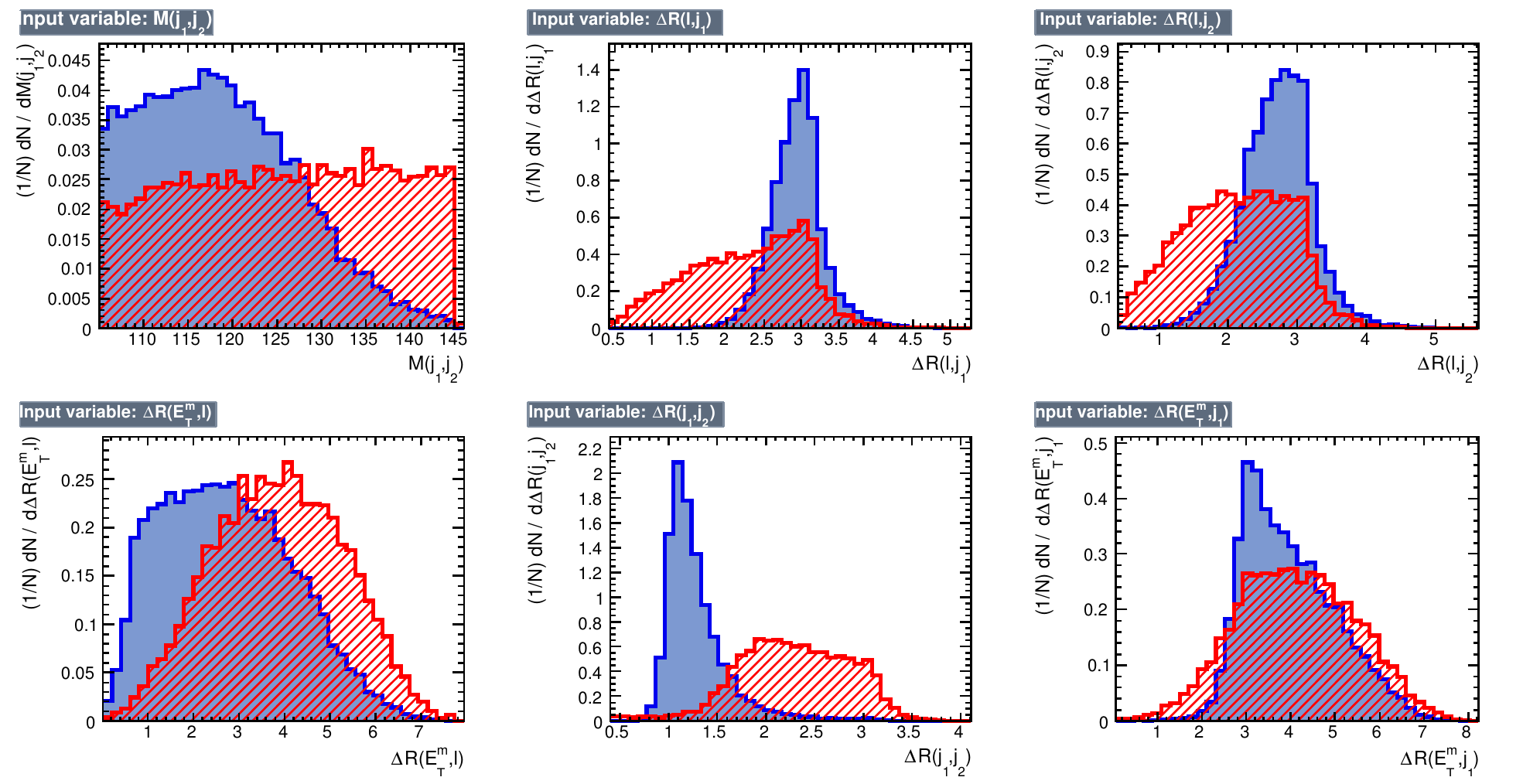}
\caption{The same as Fig.~\ref{fig:var300} but for $m_{H^\pm}=500$ GeV.}
\label{fig:var500}
\end{figure}

We choose the following twelve simple kinematic variables that are also listed in Table~\ref{tab:MVAvar} for our MVA.
\begin{itemize}
\item Transverse momenta of lepton, $p_T(\ell)$ and two leading-$p_T$ jets,
$p_T(J_1)$ and $p_T(J_2)$. 
\item Missing transverse energy $\slashed{E}_T$ and pseudorapidity of 
$\slashed{E}_T$ vector denoted by $\eta(\slashed{E}_T)$.
\item Scalar sum of transverse momenta of all visible particles denoted by $H_T$.  
\item Invariant mass of two leading-$p_T$ jets denoted by $M(J_1,J_2)$.
\item $\Delta R$ separation of $(\ell,J_1)$, $(\ell,J_2)$,
$(\slashed{E}_T,\ell)$, $(J_1,J_2)$ and $(\slashed{E}_T,J_1)$ combinations.
\end{itemize}
These variables are chosen by comparing their distributions for the signal generated for $m_{H^\pm}=300$ GeV with the total background distributions. 
They are selected from a bigger set of variables based on their discriminating power and less correlation. In Fig.~\ref{fig:var300}, we show the normalized 
distributions of these variables for the signal with $m_{H^\pm}=300$ GeV and the total background. Similar distributions for $m_{H^\pm}=500$ GeV are
shown in Fig.~\ref{fig:var500}. From these figures, one can see that each of these distributions has reasonable discriminating power between the signal and 
the background. We use these kinematic variables simultaneously in a MVA whose output shows large differences in their shapes for the signal and the background. 
One should notice that the signal distributions deviate more from the background ones as we increase $m_{H^\pm}$. Therefore, isolation of the signal from 
the background becomes easier for heavier resonances. We, therefore, tune our MVA for lower masses and use the same optimized analysis for larger masses.
\begin{table}[!h]
\centering
\begin{tabular}{cc|cc|cc|cc}
\hline \hline
Variable & Importance & Variable & Importance & Variable & Importance & Variable & Importance \\ 
\hline 
$p_T(\ell)$ & 0.095 & $\slashed{E}_T$ & 0.072 & $M(j_1,j_2)$ & 0.092 & $\Dl R(\slashed{E}_T,\ell)$ & 0.065 \\ 
$p_T(j_1)$ & 0.092 & $\eta(\slashed{E}_T)$ & 0.076 & $\Dl R(\ell,j_1)$ & 0.088 & $\Dl R(j_1,j_2)$ & 0.072 \\ 
$p_T(j_2)$  & 0.074 & $H_T$ & 0.153 & $\Dl R(\ell,j_2)$ & 0.077 & $\Dl R(\slashed{E}_T,j_1)$ & 0.044 \\ 
\hline \hline
\end{tabular} 
\caption{Input variables used for MVA (BDT algorithm) and their relative importance. These numbers are obtained for 
$m_{H^\pm}=300$ GeV for the $2b$-tag category. These numbers can vary for other choices of parameters.}
\label{tab:MVAvar}
\end{table}

In Table~\ref{tab:MVAvar}, we show the relative importance of each variable in the BDT response for $m_{H^\pm}=300$ GeV for the $2b$-tag category. 
For this particular benchmark, the $H_T$ variable has the highest relative importance of about 15\%. The greater relative importance implies that 
the corresponding variable becomes a better discriminator. Note that the relative importance of such a variable can change for other benchmarks 
and for different LHC energies that can change the shape of the distributions. It can also change due to different choices of algorithms and their 
tuning parameters.

One should always be cautious while using the BDT algorithm since it is prone to overtraining. This can happen during the training of the signal and background 
test samples due to improper choices of the tuning parameters of the BDT algorithm. One can decide whether a test sample is overtrained or not by checking 
the corresponding Kolmogorov-Smirnov (KS) probability. If it lies within the range 0.1 to 0.9, we say the sample is not overtrained. We use two statistically 
independent samples in our MVA for each benchmark mass, one for training the BDT and another for testing purposes. In our analysis, we ensure that we 
do not encounter overtraining while using the BDT by checking the corresponding KS probability.

\begin{figure}[!htbp]
\centering
\subfloat[]{\includegraphics[height=5cm,width=7cm]{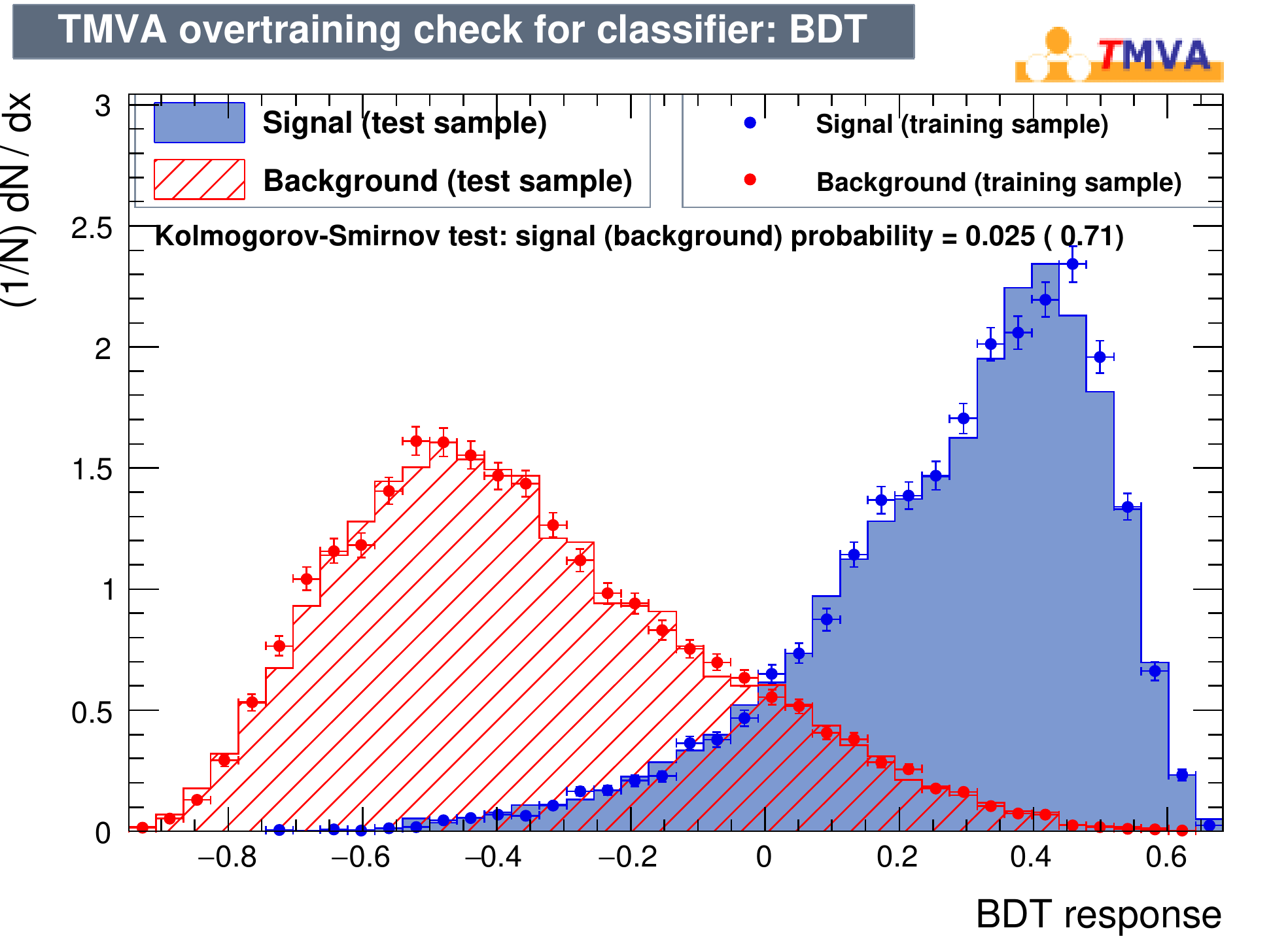}\label{fig:BDTres300}}
\subfloat[]{\includegraphics[height=5cm,width=7cm]{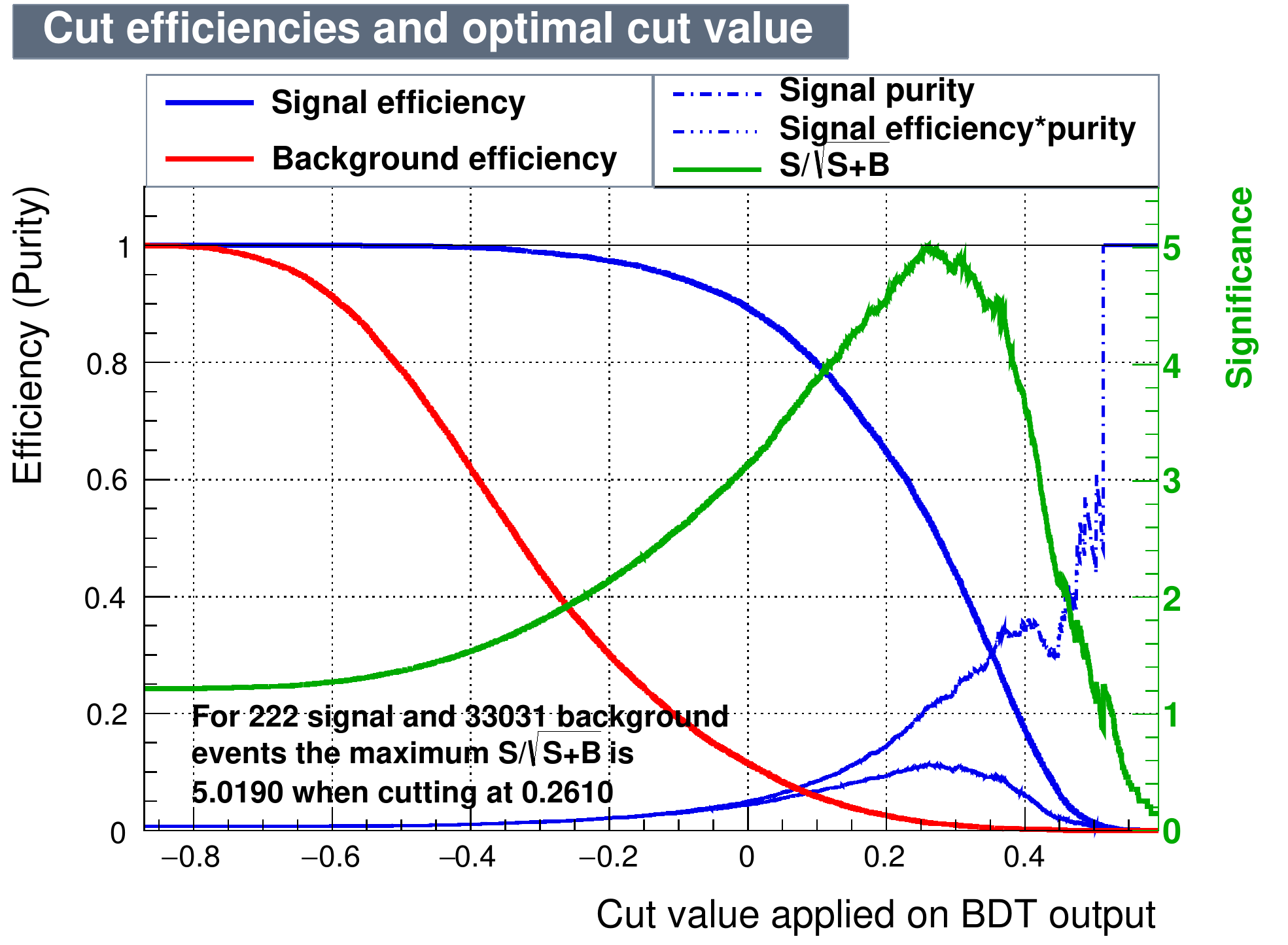}\label{fig:BDTcut300}}\\
\subfloat[]{\includegraphics[height=5cm,width=7cm]{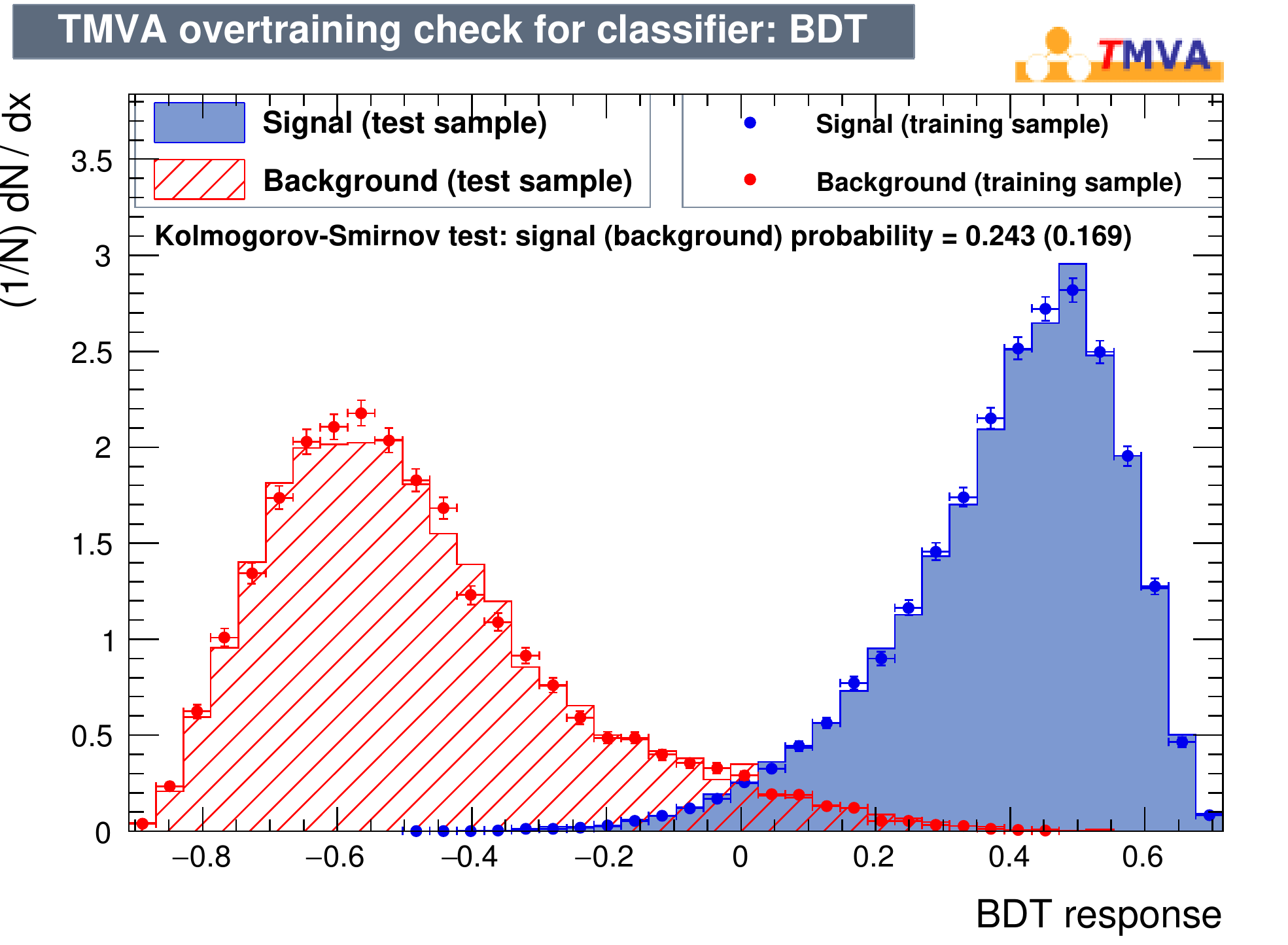}\label{fig:BDTres500}}
\subfloat[]{\includegraphics[height=5cm,width=7cm]{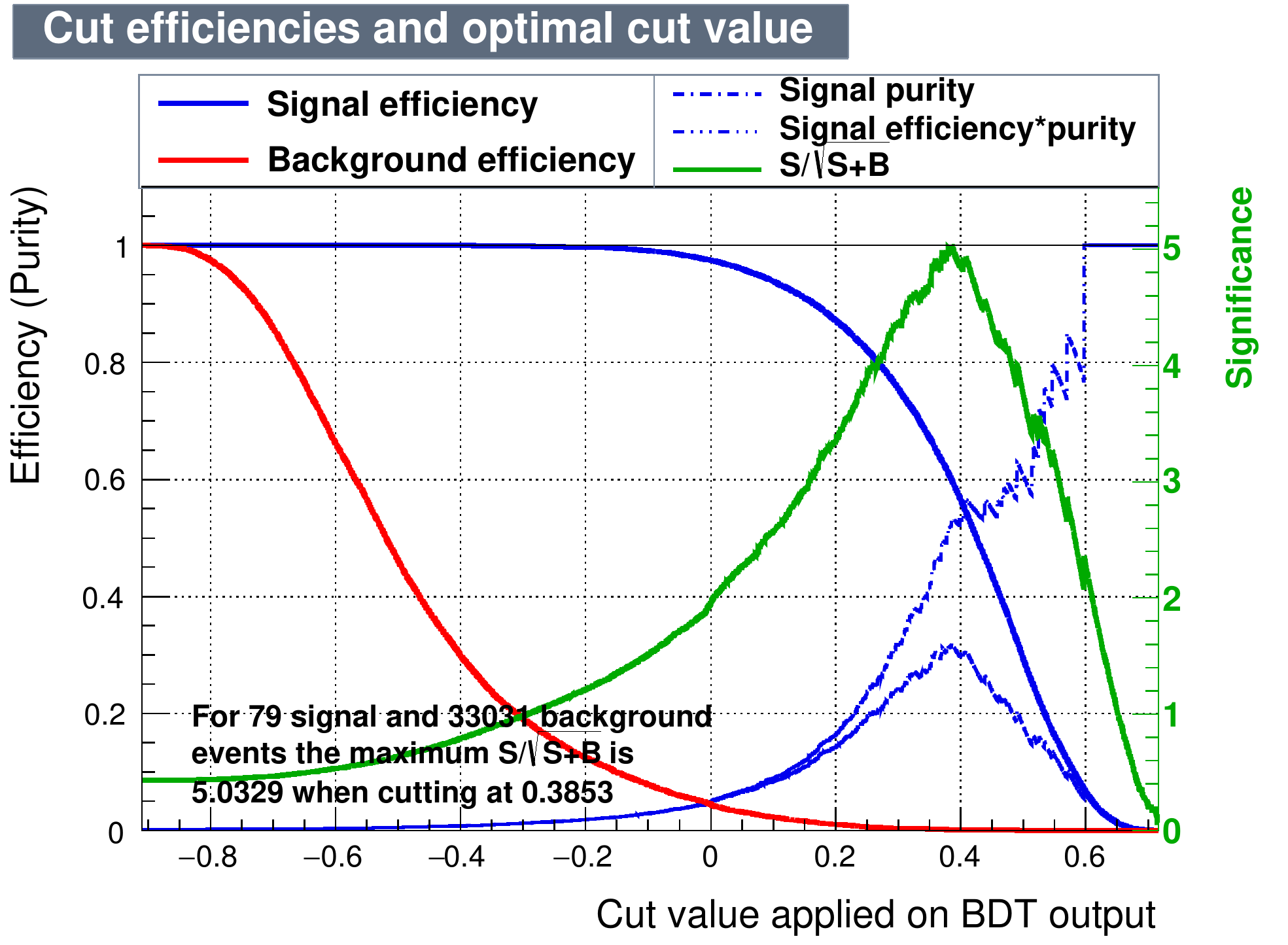}\label{fig:BDTcut500}}
\caption{(a) The BDT response for the signal and the background for $m_{H^\pm}=300$ GeV 
at the LHC ($\sqrt{s}=13$ TeV) for the 2b-tag category.
(b) The corresponding signal and background cut efficiencies and significance as functions of the BDT cut.
Discovery significance of $5\sg$ is achieved for the optimal BDT cut,
$\mathrm{BDT}_{opt}\gtrsim 0.26$. Similar figures for $m_{H^\pm}=500$ GeV
are shown in (c) and (d) where a maximum $5\sg$ significance is achieved for 
$\mathrm{BDT}_{opt}\gtrsim 0.39$.}
\label{fig:BDT}
\end{figure}

In Figs.~\ref{fig:BDTres300} and \ref{fig:BDTres500}, we display a normalized BDT output of the signal and the background for $m_{H^\pm}=300$ GeV and 
$m_{H^\pm}=500$ GeV, respectively, for the 2b-tag category at the LHC ($\sqrt{s}=13$ TeV). One can see that the BDT outputs for the signal and the background 
are well-separated, and this can improve as we go to higher $m_{H^\pm}$ values. One then applies a BDT cut i.e.~$\mathrm{BDT}_{res} > \mathcal{C}$, 
where $\mc{C}\in \lt[-1,1\rt]$ on the signal and background samples. The corresponding cut efficiencies are shown as functions of $\mc{C}$ in Fig.~\ref{fig:BDTcut300}
(Fig.~\ref{fig:BDTcut500}) for $m_{H^\pm}=300$ GeV ($m_{H^\pm}=500$ GeV). The optimal BDT cut ($\mathrm{BDT}_{opt}$) is defined for which the significance
$\mc{N}_S/\sqrt{\mc{N}_S+\mc{N}_B}$ is maximized (where $\mc{N}_S$ and $\mc{N}_B$ are the number of signal and background events, respectively, for a given
luminosity that are survived after the BDT cut). We see in Fig.~\ref{fig:BDTcut300} that if we have, at least, 222 signal events (for $\mc{L}=50$ fb$^{-1}$) before 
the BDT analysis, it is possible to achieve a maximum $5\sg$ significance for $\mathrm{BDT}_{opt}\gtrsim 0.26$. After this cut, the number of signal events is reduced to
118 from 222 but the background events are drastically reduced to 436 from 33031. In Table~\ref{tab:MVA1b}, we show $\mc{N}_S$ and $\mc{N}_B$ along with
$\mathcal{N}_S^{bc}$, the minimum number of signal events before the BDT cut that is required to achieve $5\sg$ significance, for different $m_{H^\pm}$ values 
and for the two selection categories. 
\begin{table}[!htbp]
\centering
\begin{tabular}{c|cccc|cccccc}
\hline \hline
\multicolumn{1}{c|}{$m_{H^\pm}$} & \multicolumn{4}{c|}{$1b$-tag category} & \multicolumn{4}{c}{$2b$-tag category} \\
(GeV) & $\mathcal{N}_S^{bc}$ & $\mathrm{BDT}_{opt}$ & $\mathcal{N}_{S}$ & $\mathcal{N}_{B}$  & $\mathcal{N}_S^{bc}$ & 
$\mathrm{BDT}_{opt}$ & $\mathcal{N}_{S}$ & $\mathcal{N}_{B}$\\ 
\hline 
250 & 1227 & 0.31 & 579 & 12796 & 260 & 0.23 & 151 & 758 \\ 
300 & 983 & 0.42 & 341 & 4303 & 222 & 0.26 & 118 & 436 \\ 
350 & 680 & 0.44 & 262 & 2485 & 176 & 0.29 & 99 & 295 \\ 
500 & 229 & 0.48 & 49 & 47 & 79 & 0.39 & 47 & 41 \\ 
800 & 149 & 0.43 & 55 & 66  & 60 & 0.44 & 37 & 17 \\ 
\hline \hline 
$\mc{N}_{\textrm{SM}}$ & 344173 & - & - & - & 33031 & - & - & - \\ 
\hline \hline
\end{tabular} 
\caption{The number of the SM background events ($\mc{N}_{\textrm{SM}}$) for the $1b$-tag category at the LHC ($\sqrt{s}=13$ TeV) 
with $\mc{L}=50$ fb$^{-1}$ that enters in the MVA. The minimum number of signal events that can be discovered with $5\sg$ significance 
using our MVA is denoted by $\mathcal{N}_S^{bc}$ (this is before the optimal BDT cut as shown in the third column). The signal and background 
events that survived after the optimal BDT cut are denoted by $\mc{N}_S$ and $\mc{N}_B$, respectively, and they lead to $5\sg$ significance.}
\label{tab:MVA1b}
\end{table}

\section{Discovery regions of the 3HDM parameter space}
\label{sec:translation}

The question still remains: Can the model we proposed in section~\ref{sec:3HDM} predict signals that would be visible using the presented analysis? 
In this section we find regions of the parameter space where that is the case, which shows that if limits are set by the experimental collaborations, 
the theory can be further constrained using the current experimental data. 

The first task is to match our model to the Lagrangian in Eqs.~\eqref{eq:lagkin} and \eqref{eq:lagint}. For each parameter space point, we choose the lightest 
charged scalar for the analysis. Although we concentrated our search in the parameter space region with $v_{1,2} \ll v_{3}$, as to exploit the SM-like $h_{125}$ 
state in that limit, we do not rely on the validity of the expansion in small $\xi$ in this analysis. The couplings $\kappa_{cs}$ and $\kappa_{Wh_{125}}$ are 
found in Eq.~\eqref{eq:lagWH} after a numerical calculation of the spectrum and mixing matrices. To find the discovery reach of our parameter space, we translate 
$\mc{N}_S^{bc}$ in terms of the model parameters by using the following relation 
\begin{align}
\mc{N}_S^{bc} = \sg(pp\to H^\pm \to W^\pm h_{125} \to \ell^{\pm}+\slashed{E}_T + b\bar{b})
\times \epsilon_S \times \mc{L} \,,
\end{align}
where $\sg$ is the cross section after showering and hadronization, $\epsilon_S$ is the
signal cut efficiency and $\mc{L}$ is the integrated luminosity.

For the calculation of ${\rm BR}_{Wh_{125}}$, it is important to note that although in general $H^\pm$ can decay to $W^\pm  h_{\mathrm{a,b, 125}}\, $, 
we are interested only in the decay mode involving $h_{125}$, as in our model this is the state that couples the strongest to $b \bar{b}$ 
(see Eqs.~\eqref{eq:SMHiggsY} and \eqref{eq:ExHiggsY}). 

In addition, our model must be able to pass several consistency tests in order to be phenomenologically viable, such as reproducing 
the electroweak precision measurements. The original formulation \cite{PhysRevD.46.381} for BSM contributions to the electroweak precision 
observables in terms of the $S$, $T$ and $U$ parameters assumes that the scale of new physics is $\gtrsim 1$ TeV. As our model allows for new exotic 
scalars to have masses around the electroweak scale, we must employ the more general formalism introduced in Refs.~\cite{Bamert:1994yq,BURGESS1994276} 
with an extended set of oblique parameters $S$, $T$, $U$, $V$, $W$ and $X$. These can then be used to calculate $S'$, $T'$ and $U'$ for which the standard 
$Z$-pole constraints on $S$, $T$ and $U$ apply. To compute $S'$, $T'$ and $U'$, we have applied the results in Ref.~\cite{GRIMUS200881}, 
in which $S$, $T$, $U$, $V$, $W$ and $X$ are computed for a general $N$-Higgs Doublet Model with the inclusion of arbitrary numbers 
of electrically charged and neutral $\mathrm{SU}(2)_{\mathrm{L}}$ singlets. To summarize, when scanning the model parameter space for phenomenologically interesting regions, we look for points for which the following 
constraints are satisfied:
\begin{itemize}
\item There are no tachyonic scalar masses and the scalar potential is bounded from below (the corresponding constraints on the quartic couplings 
can be found in Ref.~\cite{Keus:2013hya} taking into account that our $\lambda_{ii}$ differ by a factor two from theirs).
\item The tree-level scalar four-point amplitudes satisfy $| \mathcal{M} | < 4 \pi$.
\item The SM Higgs-like scalar has a mass no more than 5 GeV away from the observed 125 GeV value, and has a Yukawa coupling to the top quark 
satisfying $|y_{t\bar{t}h_{125}}| \in[0.9,1.1]$.
\item The exotic decays $Z \rightarrow h_{\mathrm{a,b}} A_{\mathrm{a,b}}$ are kinematically forbidden, as to not be in conflict with the precision
measurements of the $Z$ width.
\item The lightest charged Higgs has a mass in the range $[m_{H^\pm}^{(\rm{min})} , 1000 \, \rm{GeV}]$, with a different $m_{H^\pm}^{(\rm{min})}$ 
for each run (taking values $250, \,300,\, 400$ or $450$ GeV).
\item The computed values of $S'$, $T'$ and $U'$ fall within the error bars on $S$, $T$ and $U$ as reported in Ref.~\cite{Patrignani:2016xqp}.
\item The value of $\kp_{cs}^2\times {\rm BR}_{Wh_{125}}$ is at least 0.5 above the 100 fb${}^{-1}$ discovery threshold for the 1$b$-tag category 
set by the MVA.
\end{itemize}

\subsection{Scanning the parameter space}

A random scan over the parameter space of the theory is both computationally expensive and not efficient. A good alternative, without the need for 
sophisticated statistical methods but still very powerful is the use of Genetic Algorithms (GA). 

Following the guidelines set in Ref.~\cite{GAMolinaWessen}, we wrote a GA in \texttt{Mathematica} for finding the parameter points in the discovery region, 
with a fitness function taking into account all the constraints listed above and including the so-called biodiversity enhancement to explore the parameter 
space more thoroughly. 

GAs start from a randomly generated initial population, with each full cycle resulting in a new generation of candidates. The fittest parameter points 
are selected for every generation and their parameters are modified (by crossover and/or mutations) leading to a new generation. The new candidate 
points are then used in the next iteration of the GA. The GA finishes when either a maximum number of generations or a satisfactory fitness level is reached. 
We decided to build the GA relying on mutations only as it usually performs comparably to GAs including a crossover but it is simpler to implement, and 
it was stopped once a given number of valid parameter points was reached.

\subsection{Results of the GA parameter scan}

\begin{figure}[!htbp]
\centering
\subfloat[]{\includegraphics[height=5cm]{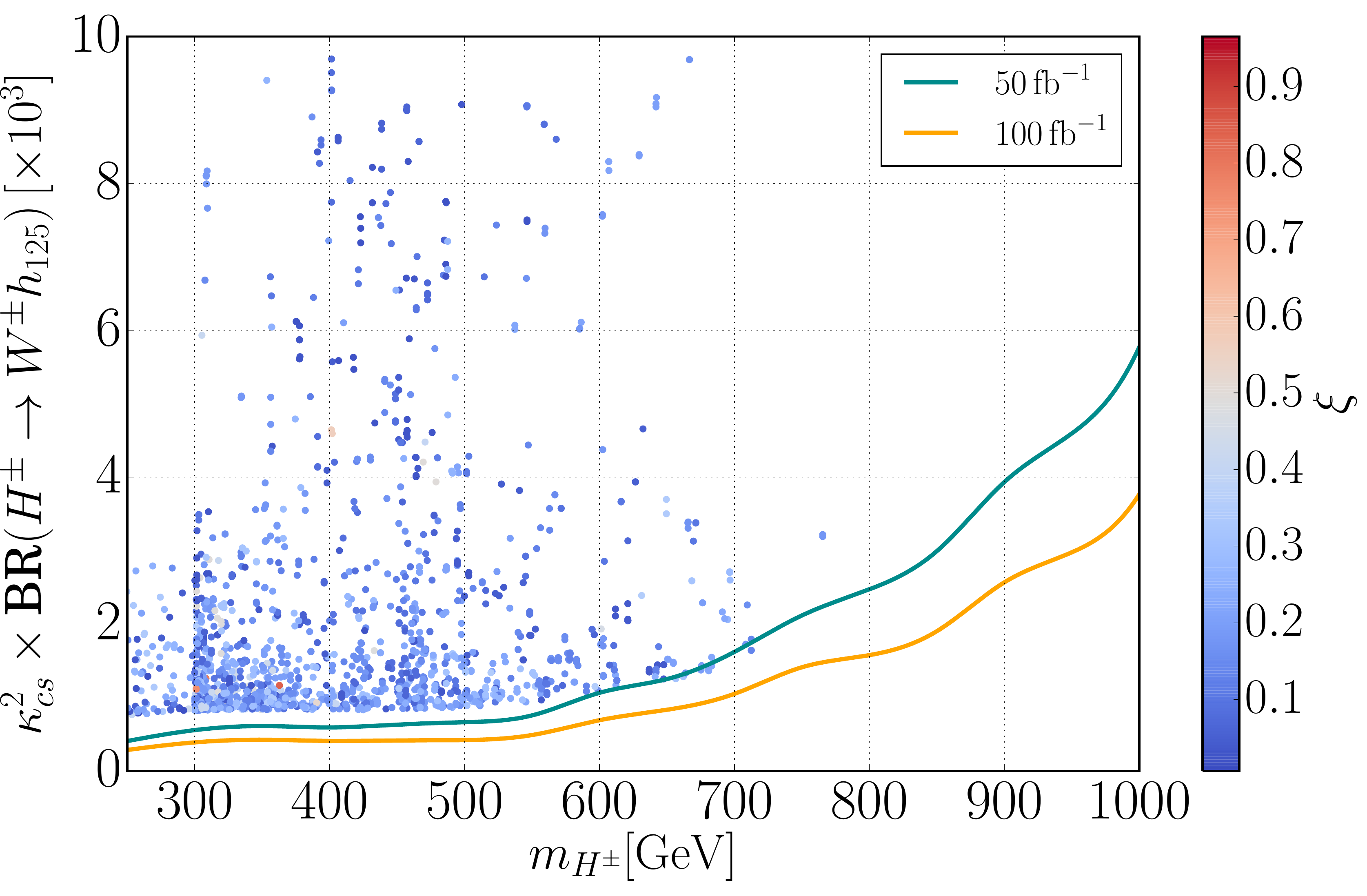}
\label{fig:5sdis1b}}
\quad
\subfloat[]{\includegraphics[height=5cm]{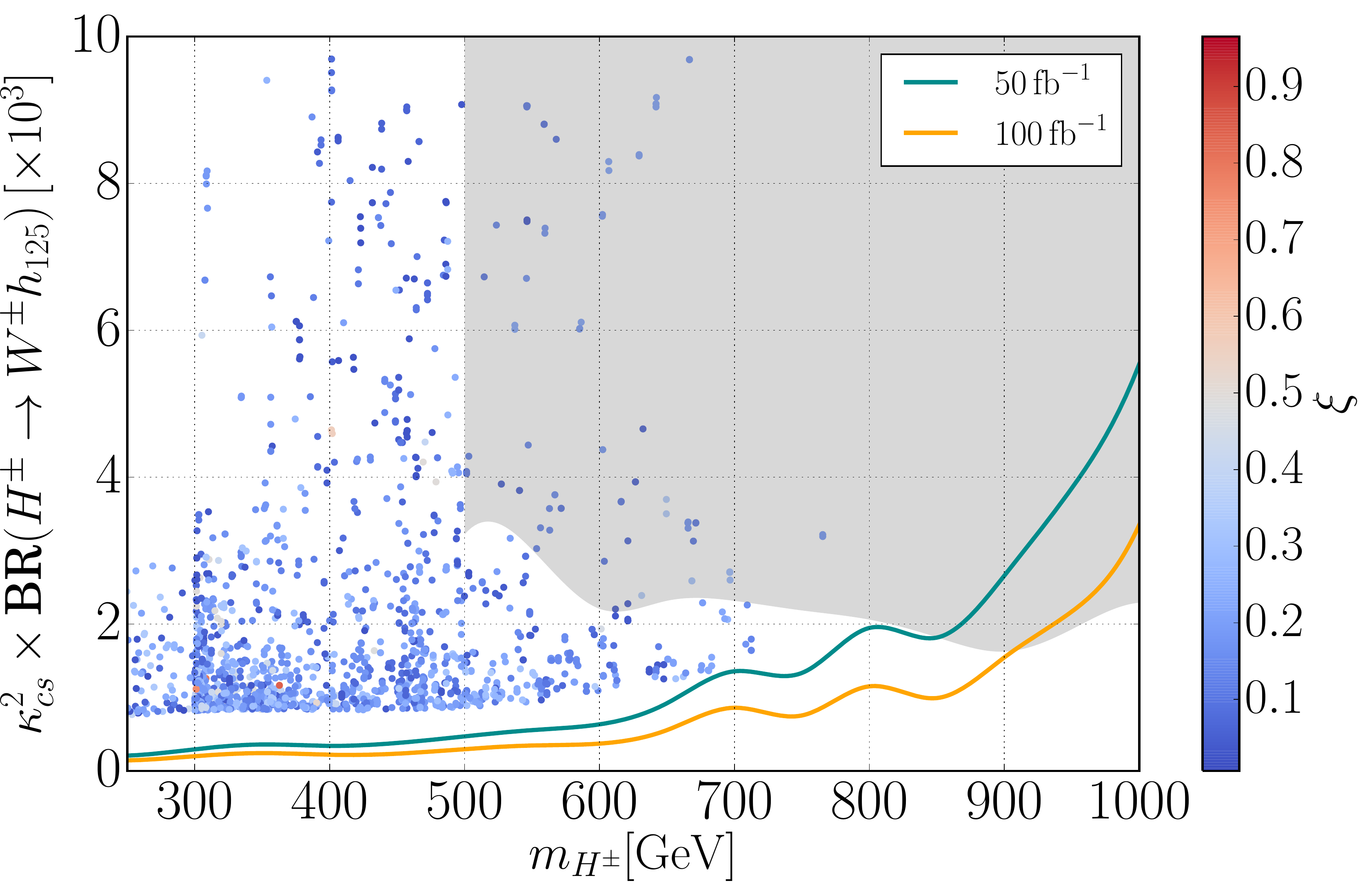}
\label{fig:5sdis2b}}
\caption{The $5\sg$ discovery contours of $\kp_{cs}^2\times {\rm BR}_{Wh_{125}}$ (scaled by $10^3$) as functions
of $m_{H^\pm}$ for $\mc{L}=50,100$ fb$^{-1}$ at the LHC ($\sqrt{s}= 13$ TeV) for (a) $1b$-tag category and 
(b) $2b$-tag category. The dots represent the parameter points resulting form the GA scan with the corresponding 
values of $\xi$ encoded in their color.}
\label{fig:5sdis}
\end{figure}

We performed five independent scans with different initial population sizes ranging from 50 to 1000, with varying mutation rates 
and different lower limits on $m_{H^\pm}$. We found 2116 parameter space points of the proposed model satisfying all the constraints within 
the discovery region of our analysis. In Figs.~\ref{fig:5sdis1b} and \ref{fig:5sdis2b}, we show the $5\sg$ discovery contours of $\kp_{cs}^2\times {\rm BR}_{Wh_{125}}$ 
corresponding to $1b$- and $2b$-tag categories, respectively, as functions of $m_{H^\pm}$ for $\mc{L}=50,100$ fb$^{-1}$ at the LHC ($\sqrt{s}= 13$ TeV). 
Here, these functions are overlayed with the corresponding values for the parameter points found by the GA scanning procedure. We find that both selection 
categories are almost equally sensitive in probing the parameter space of our model. However, the $2b$-tag category is slightly more sensitive than 
the $1b$-tag category since the background reduction is better for the former. The irregularities in the charged Higgs mass 
dependence seen in Figs.~\ref{fig:5sdis1b} and \ref{fig:5sdis2b} are due to a combination of points from scans with different lower 
limits on $m_{H^\pm}$.

As discussed before, since a $W'$ can also produce a $W h_{125}$ resonance, we compare our reach with the $W h_{125}$ resonance search data. 
In Fig.~\ref{fig:5sdis2b}, the shaded region is excluded by the ATLAS $Wh_{125}$ resonance search data~\cite{ATLAS:2017nxi} in the 
$\ell+\slashed{E}_T+b\bar{b}$ channel. To obtain this, we translate the 95\% confidence level (CL) upper limit (UL) on the cross section set by ATLAS 
in terms of our model parameters by using the following relation,
\begin{align}
\lt(\sg\times {\rm BR}\rt)_{UL}\times \epsilon_{W'}
= \sg(pp\to H^\pm)\times {\rm BR}(H^\pm\to W^\pm h_{125})\times \epsilon_{H^\pm}
\end{align}
where $\epsilon_{W'}$ and $\epsilon_{H^\pm}$ are the cut-efficiencies for the
$W'$ and $H^\pm$ respectively and they are different, in general. For simplicity, 
we assume $\epsilon_{W'}=\epsilon_{H^\pm}$ while obtaining the exclusion region
on our model parameters. For instance, for $m_{H^\pm}=800$ GeV, $\kp^2\times {\rm BR}_{Wh_{125}}\gtrsim 2\times 10^{-3}$ is excluded 
with $2\sg$ CL using $\mc{L}\approx 36$ fb$^{-1}$ data but $\kp^2\times {\rm BR}_{Wh_{125}}\lesssim 2\times 10^{-3}$ region can be discovered 
with $5\sg$ significance if we go to a higher luminosity. The exclusion region starts from $m_{H^\pm}=500$ GeV since the latest data used 
here are available for $W'$ masses above 500 GeV.  

\begin{figure}[!htbp]
\centering
\subfloat[]{\includegraphics[height=5cm]{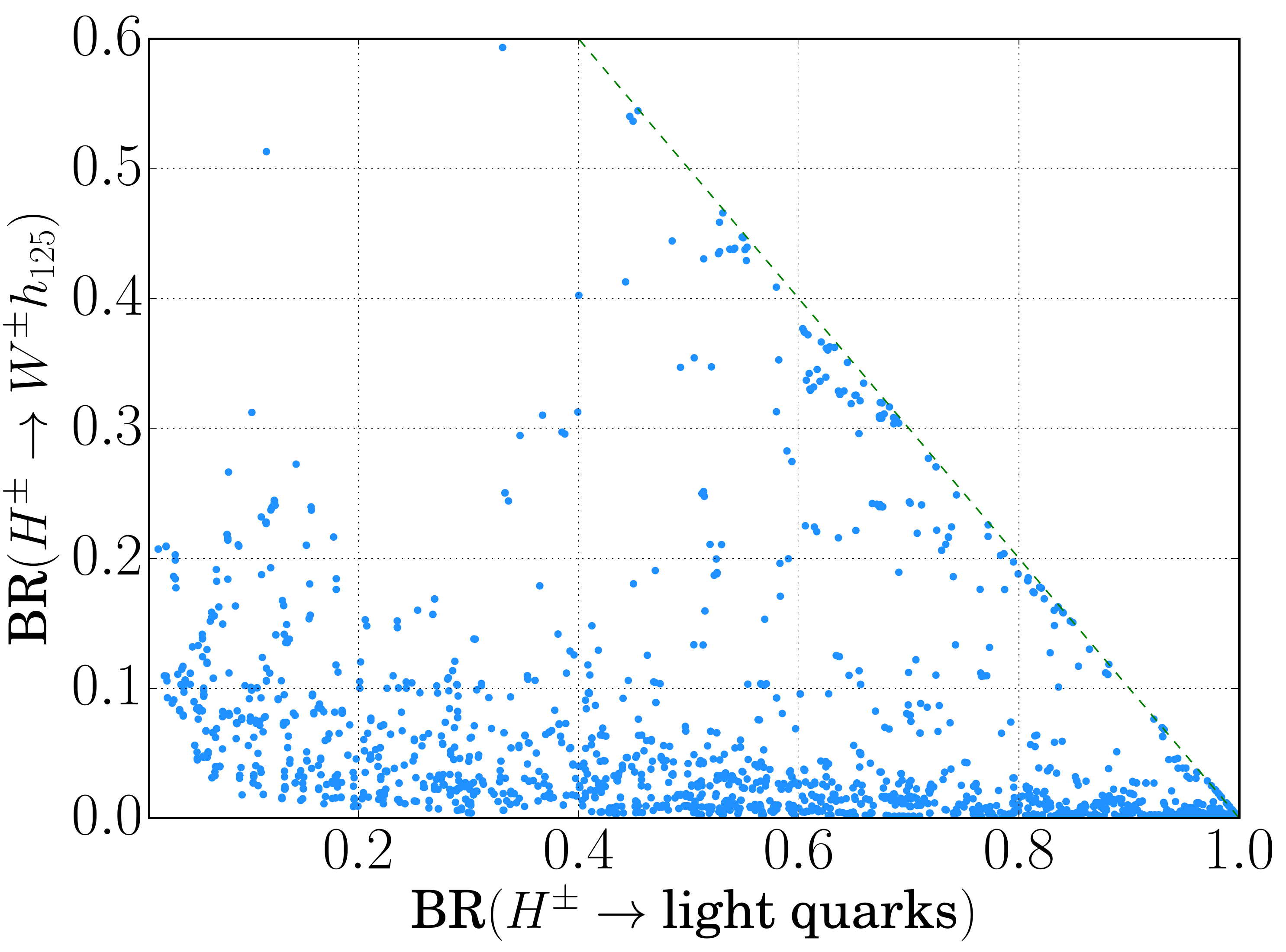}
\label{fig:Whvscs}}
\qquad
\subfloat[]{\includegraphics[height=5cm]{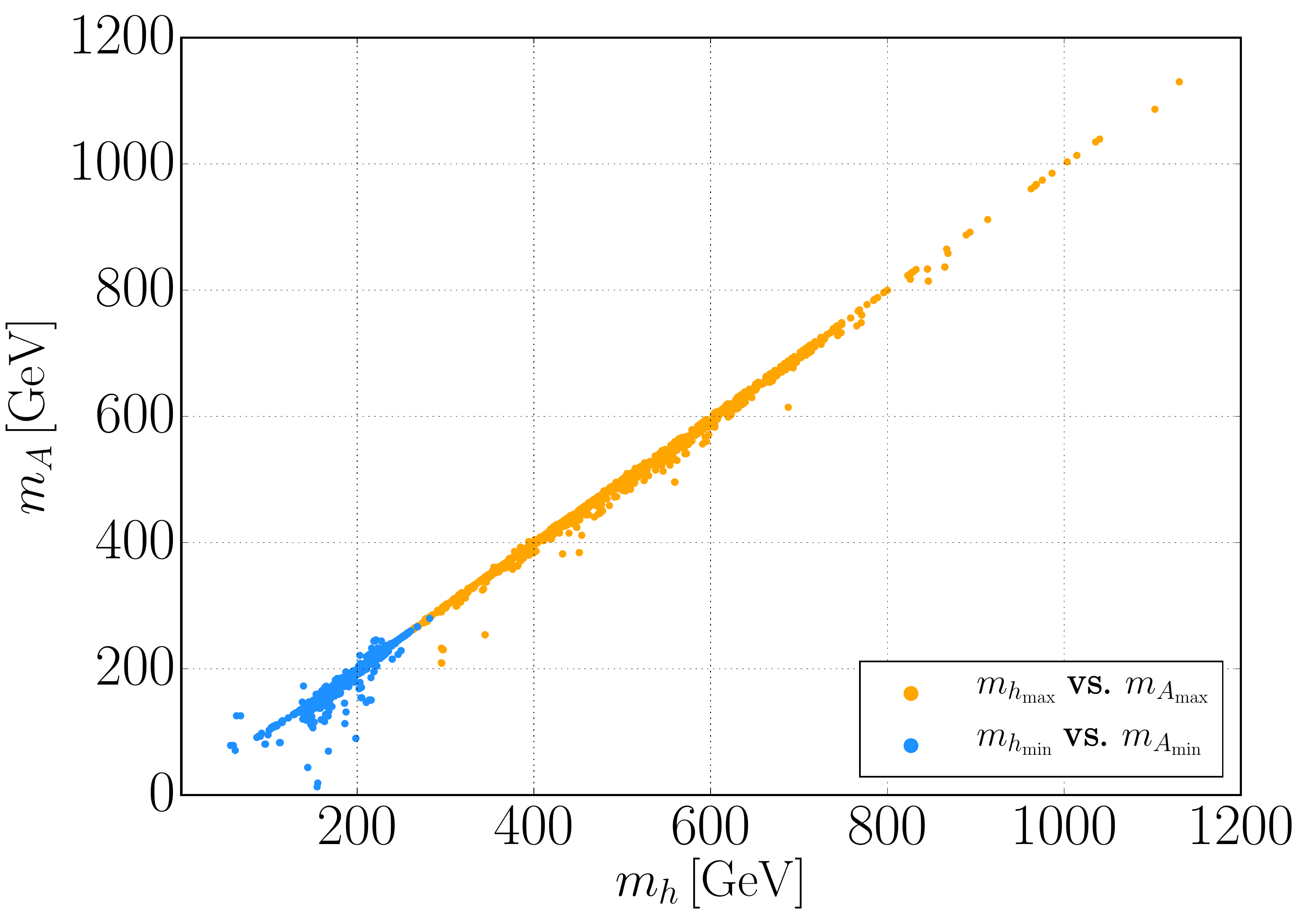}
\label{fig:MhvsMa}}
\caption{(a) ${\rm BR}(H^+ \to W^+ h_{125})$ vs. ${\rm BR}(H^+ \to \mbox{light quarks})$. Here, `light quarks' refers to 1st and 2nd generation quarks. The dashed line represents when 
${\rm BR}(H^+ \to W^+ h_{125}) + {\rm BR}(H^+ \to \mbox{light quarks})=1$, i.e.~when these two channels dominate the total decay width. 
For almost all points far away from this line, the lightest charged Higgs often decays to $W^\pm h_{\mathrm{a,b}}$. 
(b) Scalar vs.~pseudo-scalar masses for the lightest (blue) and heaviest (orange) states. The alignment of these masses 
is consistent with the $\xi\ll 1$ expansion.}
\end{figure}

Although the lightest charged scalar (identified as $H^\pm$ for the analysis) does not primarily decay into $W h_{125}$, it can still reach 
the discovery regions due to being mainly produced through $c \bar{s}$ fusion and having ${\rm BR}(W h_{125})$ comparable to the BR of the other 
decay channels. In Fig.~\ref{fig:Whvscs} we show the ${\rm BR}(H^\pm \rightarrow W^\pm h_{125})$ vs ${\rm BR}(H^+ \rightarrow \mbox{light quarks})$ for the lightest 
charged scalar, where light scalars refers to first and second generations and the dashed line represents the case when both decay modes dominate. 
For the parameter points not close to this line, the remaining decay width is mostly due to the $H^\pm \rightarrow W^{\pm} h_{\mathrm{a,b}}$ decay. For a few outlier 
points, the $H^+ \rightarrow t \bar{b}$ mode is also relevant. 
\begin{figure}
\begin{minipage}[thp]{.96\textwidth}
  \includegraphics[width=0.243\textwidth]{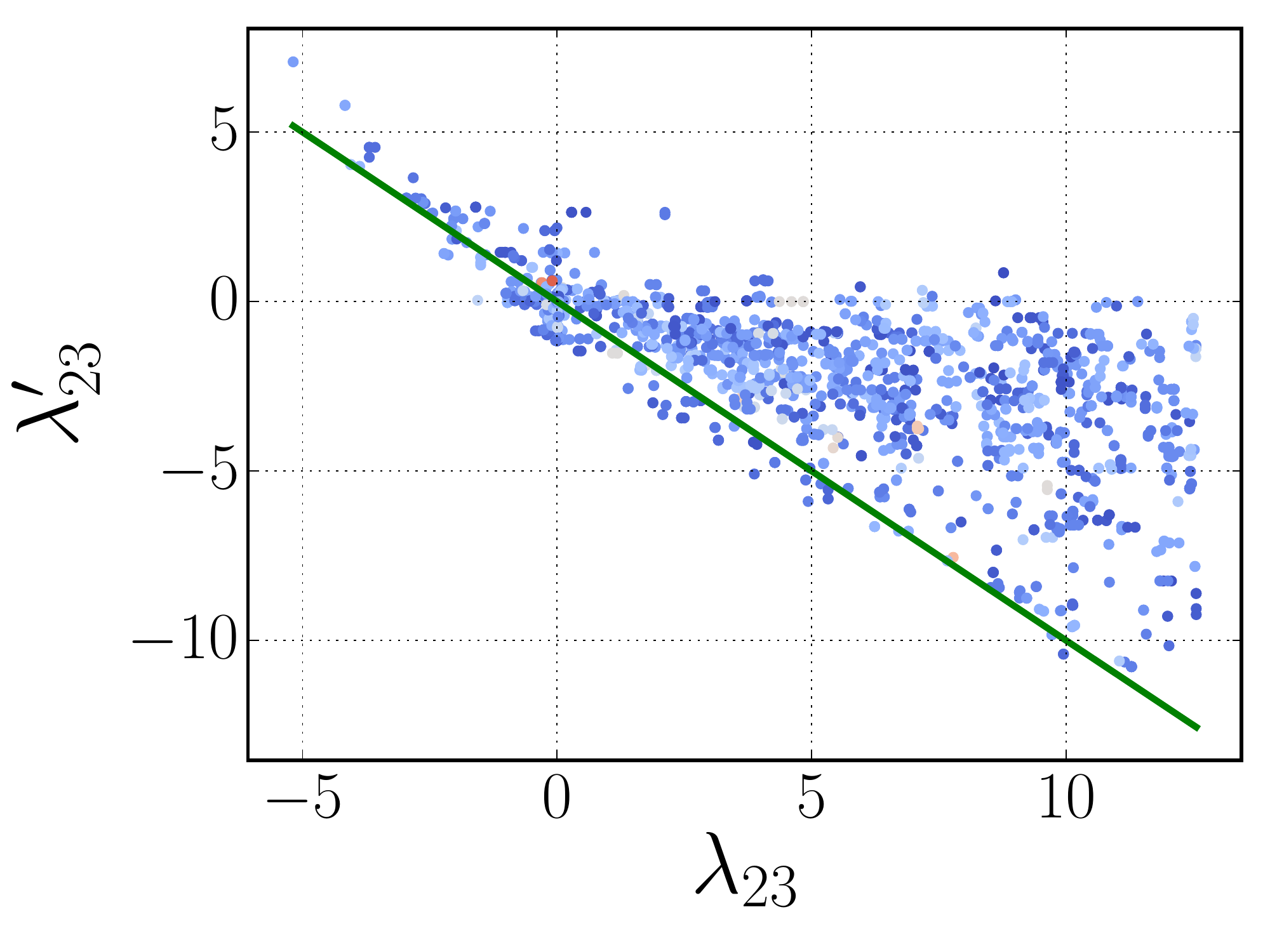}
        \includegraphics[width=0.243\textwidth]{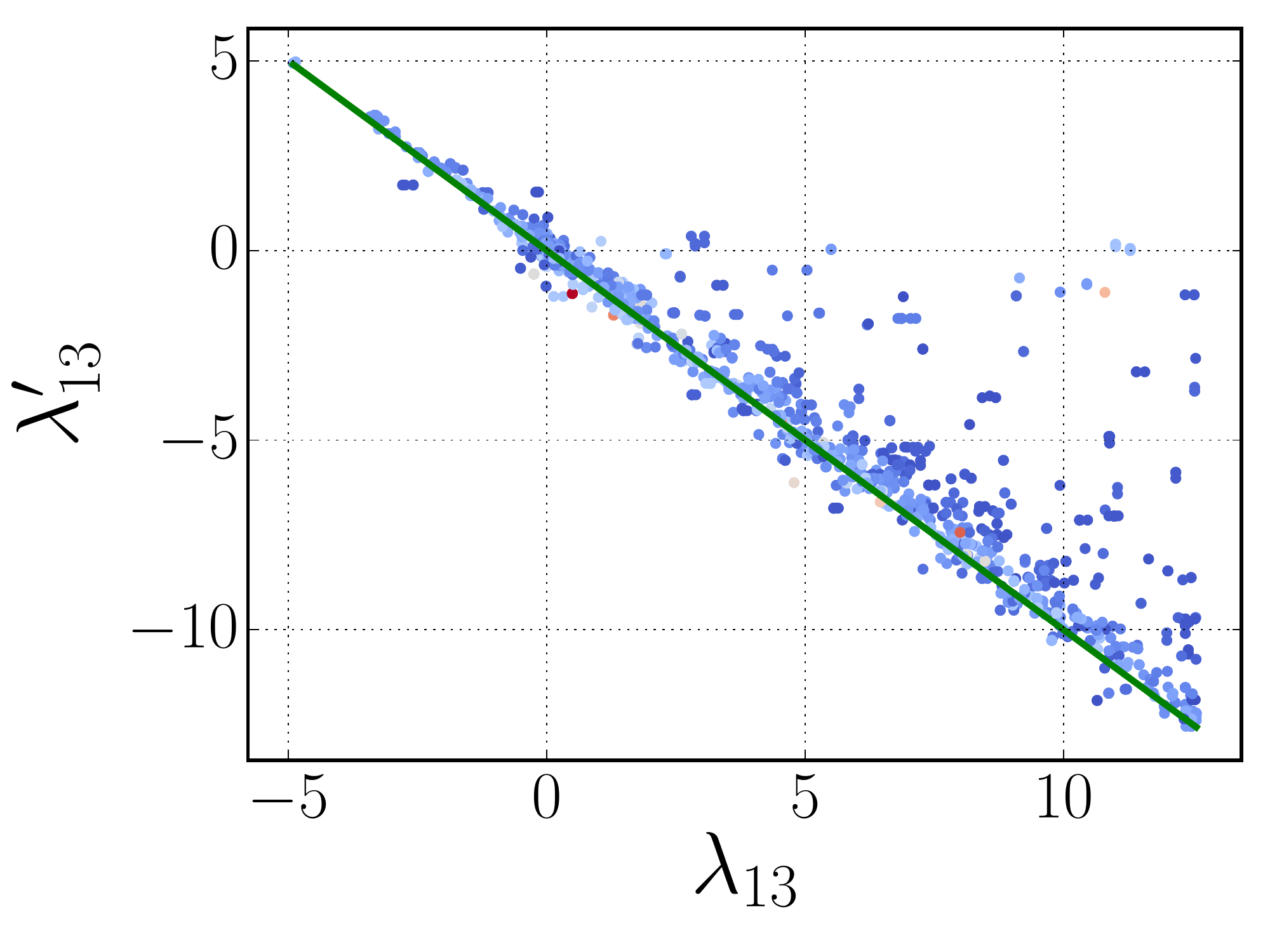}  
        \includegraphics[width=0.243\textwidth]{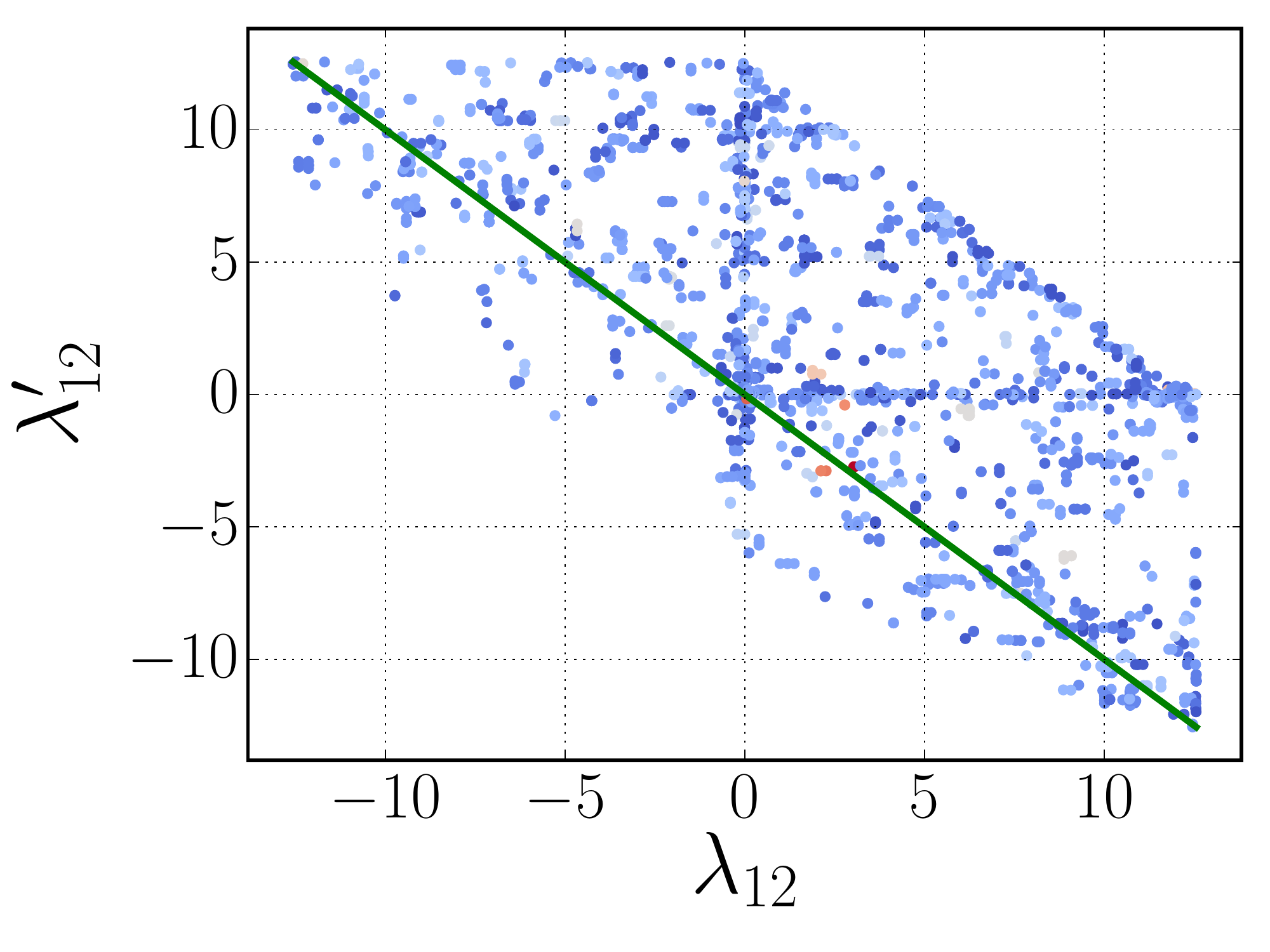}
         \includegraphics[width=0.243\textwidth]{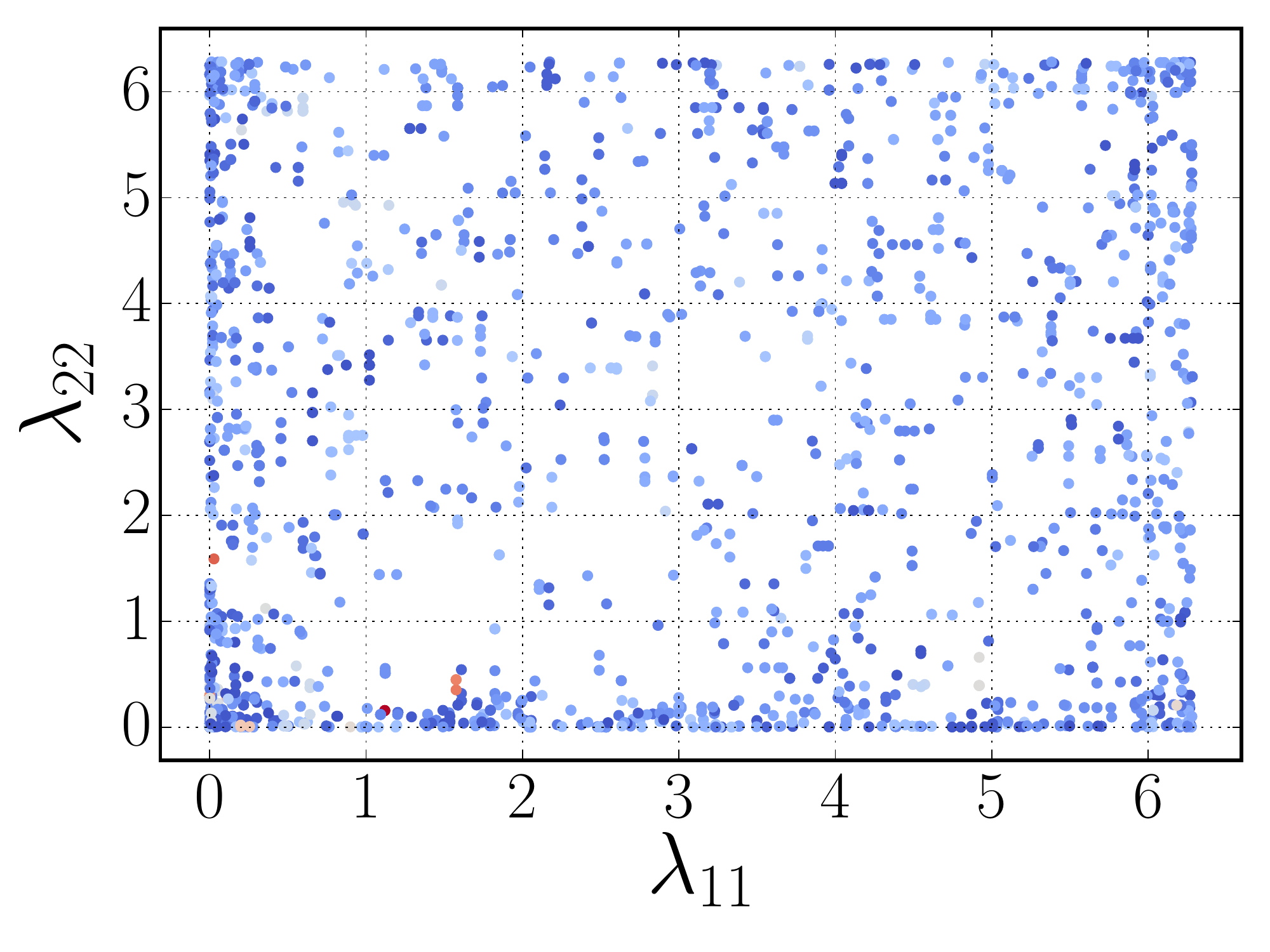}
         \includegraphics[width=0.235\textwidth]{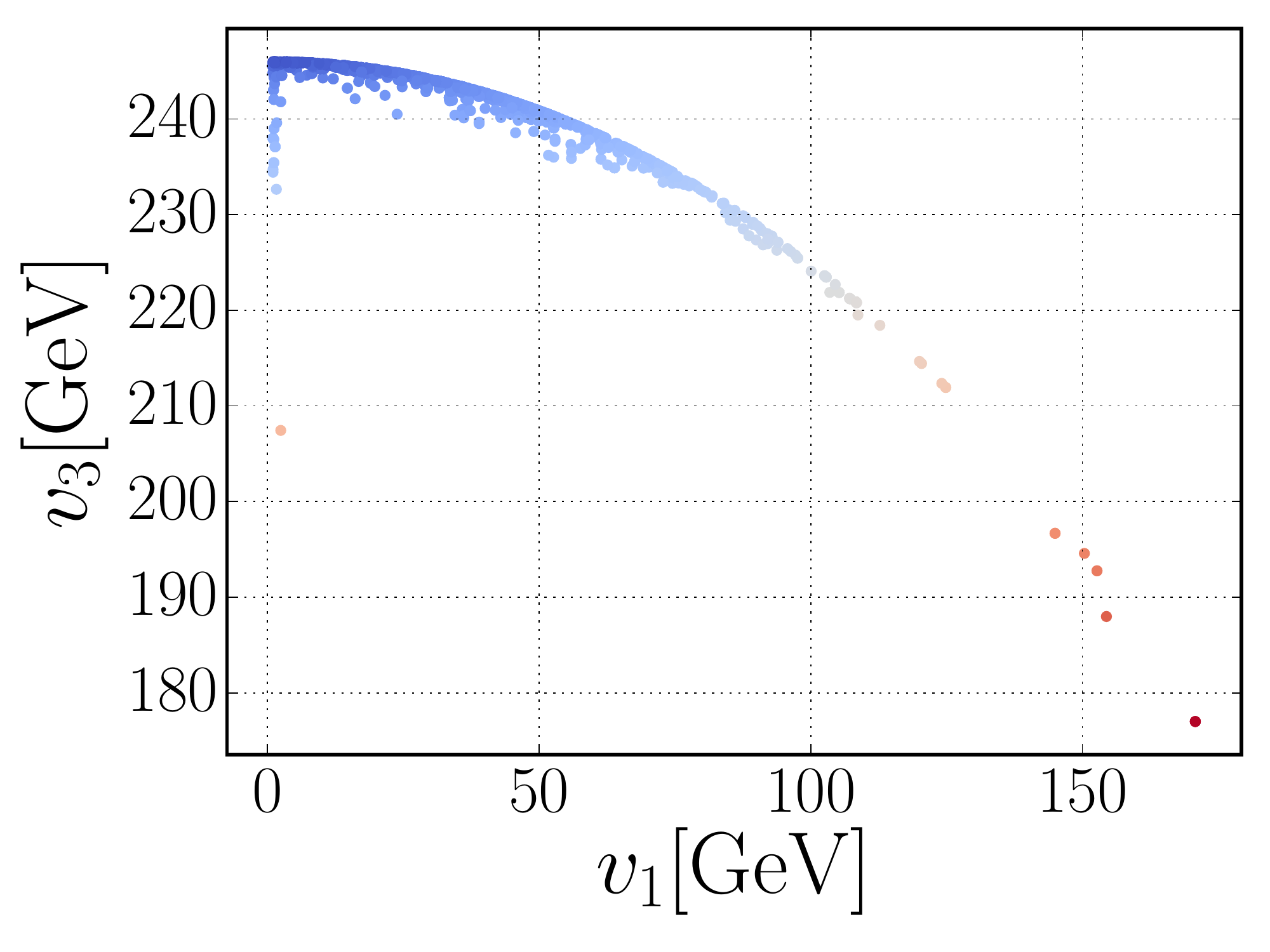}
         \includegraphics[width=0.235\textwidth]{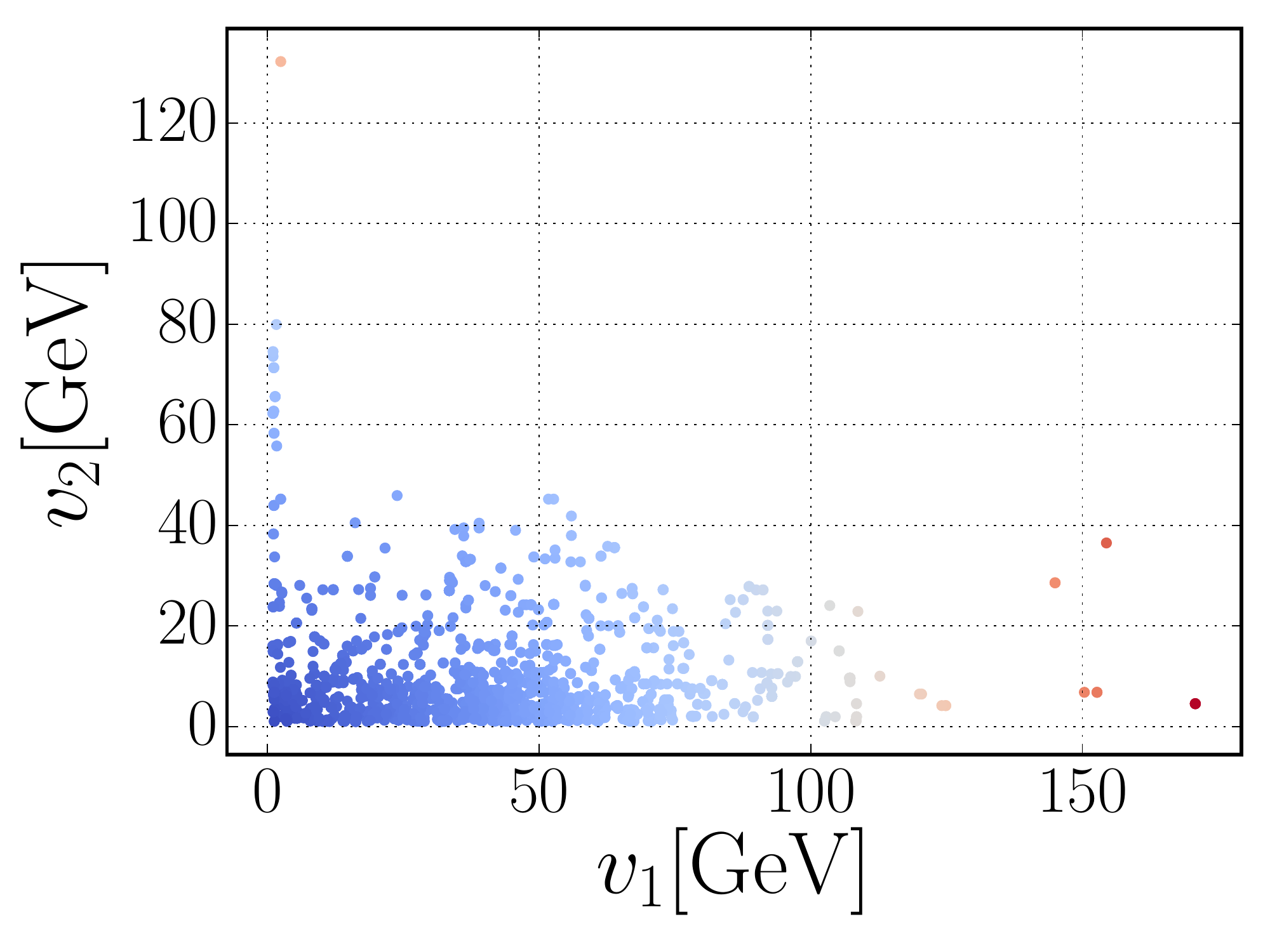}
         \includegraphics[width=0.25\textwidth]{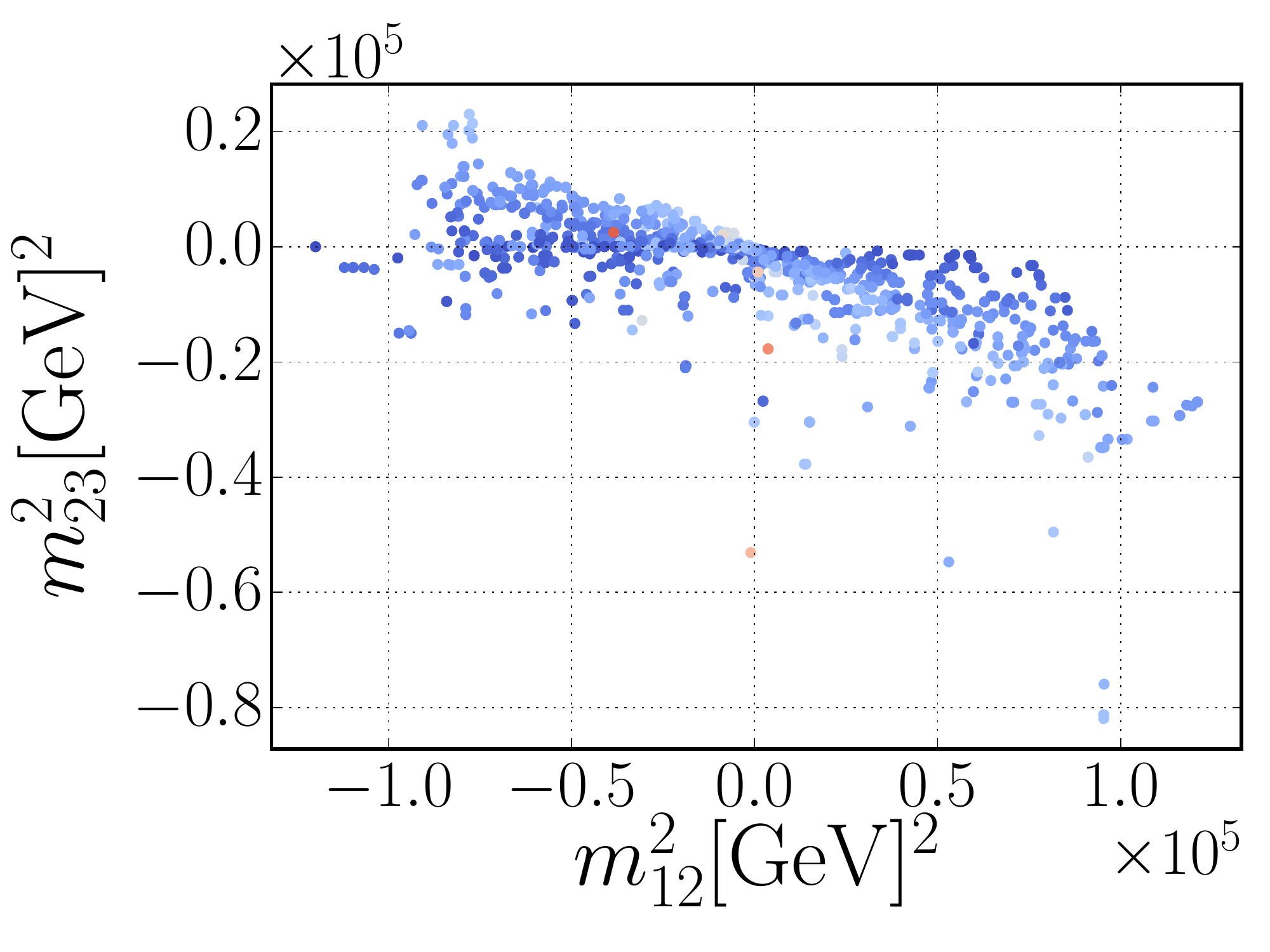}
         \includegraphics[width=0.25\textwidth]{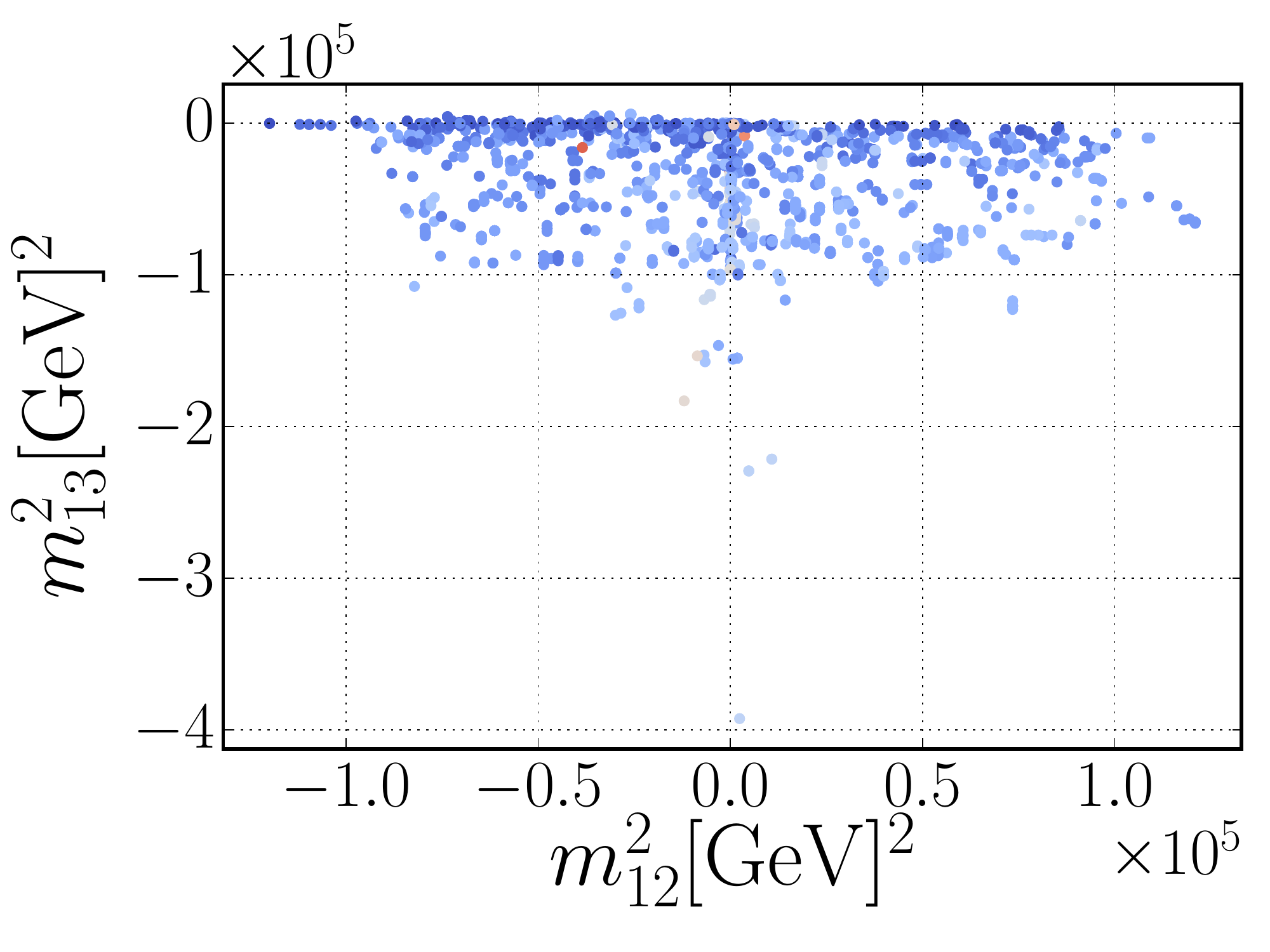}
\end{minipage}%
\begin{minipage}[thp]{.07\textwidth}
          \hspace{-11pt}
          \vspace{15pt}
          \includegraphics[height=147pt]{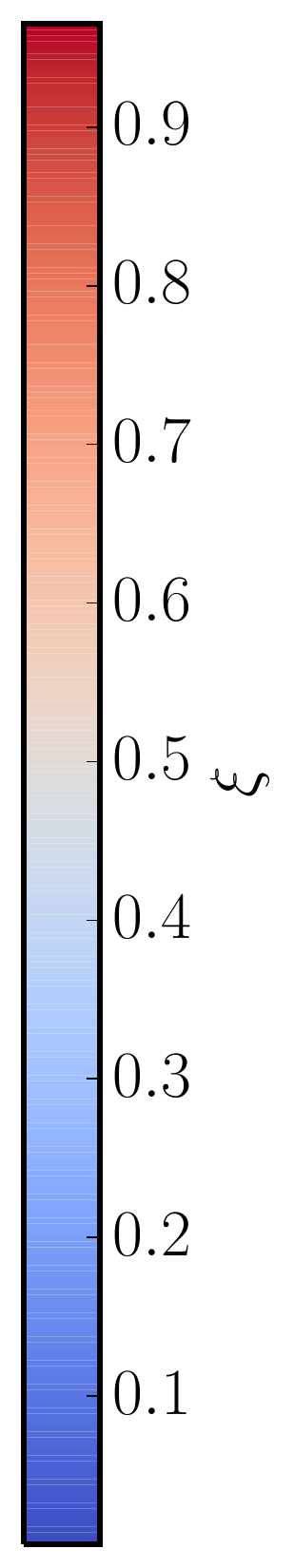}
\end{minipage}%
    \caption{Distribution of different parameters and $\xi$ values for points in the discovery region found by the Genetic Algorithm. 
    The green solid line in the first three plots indicates when $\lambda_{ij} = - \lambda^\prime_{ij}$.}
    \label{fig:pic_all}
\end{figure}

It is worth noting that although the GA did not rely on the validity of the $\xi\ll 1$ expansion, it often found points where that is the case. 
Although the initial populations had a hierarchy in the VEVs of the scalar fields ($v_{1,2} \ll v_3$), the GA had no inherent constraints stopping it from exploring the 
regions without it. Notably, the majority of the found points did show that feature and therefore a diminished hierarchy in the Yukawa 
couplings of the quark sector together with all the features described in section~\ref{sec:3HDM}. That can also be seen in Figs.~\ref{fig:5sdis1b}--\ref{fig:5sdis2b} 
where we have indicated the specific values of $\xi$ for the valid parameter points. We have checked the difference between the $\mathcal{O}(\xi)$ 
expressions for masses and the full numerical calculation, and find that for a vast majority of the valid parameter points the $\xi\ll 1$ expansion is reliable.

In Fig.~\ref{fig:pic_all}, we show distributions of quartic couplings, values of $\xi$ and mass parameters in the scalar potential, as well as the Higgs VEVs, for points 
in the discovery region found by the GA. Indeed, the typical values of $v_{1,2}$ are likely to be below 100 GeV, with $v_2$ extending over a larger
domain than $v_1$, while $v_3$ values are mostly concentrated close to the maximal 246 GeV limit. There is still a small number of valid points incompatible 
with the $\xi$-expansion due to the smallness of one of $m^2_{ij}$. Such points would still be in the discovery region, although without the features that 
assume $\xi \ll 1$. As can be seen from Fig.~\ref{fig:MhvsMa}, the masses of the exotic scalars and pseudo-scalars tend to align as predicted by the $\xi\ll 1$ 
expansion (see Eq.~\eqref{eq:ScalarMasses}). 

\section{Summary and conclusions} 
\label{sec:sumconclu}

We have, in this article, introduced a class of 3HDMs with a global $\U{X} \times \U{Z}$ family symmetry that is softly broken by bi-linear terms 
in the scalar potential. We have shown how to assign the $X$ and $Z$ charges of the quarks such that no tree-level FCNCs are present, while 
enforcing the Cabibbo-like structure of $V_{\mathrm{CKM}}$. We described how a mixing with the third quark family can be induced from dim-6 
operators, which would explain the smallness of the corresponding entries in  $V_{\mathrm{CKM}}$. Moreover, we showed that a hierarchy 
in the VEVs of the three Higgs doublets, $v_{1,2} \ll v_3$, leads to a heavy third quark family without the need for a strong hierarchy in 
the Yukawa couplings (contrary to what happens in the SM where e.g.~$y_{\mathrm{up}}/y_{\mathrm{top}} \sim 10^{-5}$). The same hierarchy 
has been exploited to derive simple closed expressions for the scalar masses and mixing matrices by expansions in the small parameter 
$\xi \equiv \sqrt{v_1^2+v_2^2}/v_3\ll 1$.

A generic prediction of the model is that the new scalars $h_{\mathrm{a,b}}$, $A_{\mathrm{a,b}}$ and $H_{\mathrm{a,b}}^{\pm}$ are likely to couple 
strongly to the $s$ and $c$ quarks, yielding different signatures in colliders at variance with the standard searches focusing on the third quark family. 
As an example, we studied collider phenomenology of the lightest charged Higgs when its mass is in the 250 -- 1000 GeV range, under the assumption 
that the other charged Higgs is sufficiently heavy to be dropped out of the analysis. In that case, the lighter charged Higgs would be resonantly produced 
through a $c\bar{s}$ fusion, and, for certain regions of the parameter space, subsequently decay to $W h_{125}$. All other decay channels are assumed 
to only contribute to its total width.

We particularly focused on one of the possible channels -- the $c\bar s\to H^+\to W^+\,h_{125}$ channel, which has not been explored before 
in the context of heavier charged Higgs searches. This channel is specific to our class of 3HDMs and is particularly sensitive to the sub-TeV charged Higgs mass
and small-$\xi$ regions. We showed that this unconventional channel, when combined with the power of a multivariate analysis, leads to good 
signal-to-background ratios even for masses below 500 GeV and thus can be used to probe models with that particular feature at the LHC. 
We employed a model independent formulation so that our approach can be applied to any model which predicts a sufficiently large cross section for 
the $c\bar{s} \to H^+ \to W^+\,h_{125}$ process to be observed in the future LHC runs. Our analysis can also be applied to improve sensitivity 
for $W'$ searches especially for the sub-TeV masses.

We then used a genetic algorithm to find parameter space points in our 3HDM which would yield signals with $>5 \sigma$ significance, while still satisfying 
the standard phenomenological constraints. Although the scan did not rely on $\xi \ll 1$, a vast majority of the points were consistent with that limit and 
thus showed all the features mentioned above and described in section~\ref{sec:3HDM}. This shows that the described unconventional search strategy 
can effectively probe realistic multi-Higgs theories with the current LHC data, and so we think it should be seriously considered by our experimental colleagues.

\section*{Acknowledgments}

The authors would like to thank Johan Rathsman for fruitful discussions.
The work of T.M. is supported by the Swedish Research Council under contract 621-2011-5107 and 2015-04814 and 
the Carl Trygger Foundation under contract CTS-14:206.~J.~E.~C.-M. was partially supported by Lund University.~R.P.~and 
J.W.~were partially supported by the Swedish Research Council, contract numbers 621-2013-428 and 2016-05996.~R.P.~was 
also partially supported by CONICYT grant PIA ACT1406.

\bibliography{reference}{}
\bibliographystyle{unsrt}

\end{document}